%% file: BoostingCopulaTimeToEvent_MainManuscript.tex
\documentclass[a4paper,12pt]{article}
\usepackage{graphicx} % Required for inserting images
\usepackage{amsmath, amsthm}
 \usepackage{array}
 \usepackage{float}
\usepackage{amssymb}
\usepackage{amsfonts}
\usepackage{dsfont}
\usepackage{booktabs}
\usepackage{graphicx}
\usepackage{multirow}
\usepackage[table, svgnames, dvipsnames]{xcolor}
\usepackage{natbib}
\usepackage[english]{babel}
\usepackage[nottoc]{tocbibind}
\usepackage{rotating}
\usepackage{pifont}% http://ctan.org/pkg/pifont
\usepackage{textcomp, gensymb}
\usepackage{hyperref}
\usepackage{afterpage}
\usepackage{setspace}
\usepackage[font=small]{caption}

\usepackage{algorithm}
\usepackage{algpseudocode}
\usepackage{adjustbox}
\usepackage{setspace}
\usepackage{geometry}
\usepackage{comment}
\input{defs.tex}

\title{Boosting Distributional Copula Regression for Bivariate Right-Censored Time-to-Event Data}
\author{Guillermo Brise\~no Sanchez$^1$, Nadja Klein$^1$, 
Andreas Groll$^2$ and Andreas Mayr$^3$ }
\date{%
    $^1$Methods for Big Data, Scientific Computing Center, Karlsruhe Institute of Technology, Karlsruhe, Germany,\\[2ex]%
    $^2$Statistical Methods for Big Data, Department of Statistics, TU Dortmund University, Dortmund, Germany,\\[2ex]%
    $^3$Department of Medical Biometry und Statistics, Philipps University of Marburg, Marburg, Germany.\\[2ex]%
    \today
}

\begin{document}

\maketitle
\thispagestyle{empty} 

%\vspace{-1cm}
\begin{abstract}
We propose a highly flexible distributional copula regression model for bivariate time-to-event data in the presence of right-censoring. The joint survival function of the response is constructed using parametric copulas, allowing for a separate specification of the dependence structure between the time-to-event outcome variables and their respective marginal survival distributions. The latter are specified using well-known parametric distributions such as the log-Normal, log-Logistic (proportional odds model), or Weibull (proportional hazards model) distributions. Hence, the marginal univariate event times can be specified as parametric (also known as \textit{Accelerated Failure Time}, AFT) models. Embedding our model into the class of generalized additive models for location, scale and shape, possibly all distribution parameters of the joint survival function can depend on covariates. We develop a component-wise gradient-based boosting algorithm for estimation. This way, our approach is able to conduct data-driven variable selection. To the best of our knowledge, this is the first implementation of multivariate AFT models via distributional copula regression with automatic variable selection via statistical boosting.  A special merit of our approach is that it works for high-dimensional $(p \gg n)$ settings. We illustrate the practical potential of our method on a high-dimensional application related to semi-competing risks responses in ovarian cancer. All of our methods are implemented in the open source  statistical software \textsf{R} as add-on functions of the package gamboostLSS.
\end{abstract}

\vspace{1em}


\pagebreak
\setcounter{page}{1} 

\onehalfspacing

\section{Introduction}\label{SectionIntro}

Advancements in molecular medicine, genetics and digital transformation of healthcare have facilitated the collection of large-scale data structures related to individual patients. Some prominent examples are Genome-Wide Association Studies \citep[GWAS;][]{Uffelmann2021} and The Cancer Genome Atlas Program \cite[TCGA;][]{TCGA}. Various techniques have been developed to analyse such ``omics'' data in a concise, scalable manner, while at the same time preserving the interpretability of the results. An important challenge when facing a vast amount of potentially influencing factors is to find a subset of such factors that has the most impact on the outcome of interest. For exploratory analyses, taking into account the entire information simultaneously instead of performing multiple univariate analyses that ignore the remaining variables in the data is of great importance. Individual analysis of the potential influencing factors without consideration of the remainder could lead to estimation bias or falsely informative selected variables. Therefore, the aforementioned \textit{variable selection} procedure should have as least input from an analyst as possible and instead rely on data-driven techniques. 

Compared to classical continuous or binary endpoints, time-to-event data are typically incomplete or \textit{censored} for individual where the event of interest was not observed. Conducting statistical analysis without taking censoring into account leads to bias in the estimation, which could result in incorrect treatment, diagnosis and prognosis. Time-to-event analyses or ``survival analyses''  \citep{KleMoe2003}  explicitly account for censored observations, see \cite{BeiIliPap2024} for a review focused on clinical applications. When analysing univariate censored event-time responses in a regression context, the Cox proportional hazards model \citep{Cox1972} is one of the most popular methods, although the interpretation of hazards remains challenging 
\citep{heller2024simple, beyersmann2024discussion}.

A wide range of tools for analysing univariate time-to-event responses accompanied by a large amount of covariate information are available. One commonly used technique to navigate large data structures with high-dimensional covariate information is based on univariate modelling paired with hypothesis testing \citep{ChoTur2020, JenKuoSto2002}. That is, the response is modelled as function of one covariate, and after carrying out all of the univariate combinations the p-values obtained from the statistical tests are sorted in ascending order. Afterwards, a subset that includes the ``most significant'' variables is chosen. In the context of genomics, where gene expression data is overwhelmingly large relative to the number of observations, following the aforementioned approach may lead to poor results \citep{LoCheZhe2015}. More sophisticated variable selection approaches such as the LASSO have been adapted to the Cox model \citep{Tib1997} as well as Accelerated Failure Time (AFT) or parametric survival models, see e.g.\ \cite{ParTagSe2024}. More recently, ``black-box'' or less interpretable methods have also been proposed by \cite{IshKogUda2011}, \cite{NorLiWen2024}, and \cite{WanLi2024}, to name a few, and \cite{SalLi2023} for a review. The main limitation of the aforementioned contributions is  their restriction to univariate time-to-event responses. 

While a broad literature on multivariate time-to-event analysis exists,  variable selection in these models remains somewhat unaddressed. Current proposed approaches do not scale to higher dimensions of covariate information or have not adopted a data-driven approach to variable selection. \cite{Marra2019_CLBMODELS_JASA} introduced a  flexible class of bivariate time-to-event models using parametric copulas. In their approach, the marginal survival functions are modelled semi-parametrically using additive regression techniques and smooth functions of time. \cite{SunDing2019} proposed a copula-based model for time-to-event analysis as well, albeit their implementation is tailored towards interval-censored responses, marginal distributions being of the same family, and the dependence between the event times cannot depend on  covariates. A copula-based model for correlated event times was proposed by \cite{EmuNak2017}. However, their approach resorts to ``Cox-type'' specifications of the marginal survival functions and is also restricted to a constant dependence parameter. Moreover, \cite{EmuNak2018} extended their proposed model to (indirectly) account for high-dimensional covariates using  a ``composite covariate'' \citep{Tukey1993}, where a linear combination of coefficients and covariates summarises the high dimensional covariate vector to a scalar variable or index. This new scalar variable is used as proxy for the original high-dimensional covariate information.

In summary,  limitations of the currently available methods for time-to-event analysis may be assigned to three categories: (1) The approaches offer solutions for high-dimensional covariates, but are restricted to univariate time-to-event responses. (2) The approaches are able to model multivariate event times, but restrictions exist regarding the flexibility of the marginal survival functions, dependence structure, or covariate effects. (3) The methods are able to handle multivariate responses, but do not scale to high-dimensional covariates or rely on heuristics or non-interpretable techniques to tackle this issue. 

We aim to address these gaps by proposing a flexible approach that  allows to account for different types of covariate effects in a copula-based multivariate time-to-event  model. Furthermore, our proposal allows for scalable, data-driven variable selection via estimation through statistical boosting \citep{BuehlmannHothornBOOSTING}. Boosting has been explored previously in a univariate time-to-event context using different modelling approaches. For example, \cite{BinAllSch2009} applied boosting to high-dimensional competing risks data. \cite{HeLiZhu2016} applied it for false discovery control, whereas \cite{MayHofSch2016} focused on optimising the concordance index.
More recently, \cite{MorHe2020} released a package for boosting stratified Cox proportional hazards models. In terms of multivariate responses, \cite{GriGroBer2021} proposed a boosting algorithm for variable allocation and selection in the context of joint models for longitudinal and survival data, see  \cite{JointModelsBooks} for more on this model class. Lastly, the alternative modelling paradigm of ``first-hitting-time'' was combined with boosting by \cite{DeBinSti2023}. Our proposed statistical modelling framework allows to construct  flexible parametric joint survival functions based on the copula approach. A main advantage is to potentially model all parameters of the joint survival function as functions of covariates using structured additive predictors \citep{Woo2017}. This in principle gives directly interpretable models. However, because we allow all distribution parameters to depend on covariates,  scalable and data-driven variable selection without any input from the analyst is highly desirable. To achieve this goal,  we suggest estimation via statistical boosting building on the work of \cite{HansBoostDistReg} and \cite{BriKle2024}. Compared to these authors, we thereby provide boosting methodology and software implementation for distributional copula regression by allowing the responses to be subject to independent right-censoring. To the best of our knowledge, this is the only publicly available software implementation that allows to fit bivariate time-to-event models which combines a wide range of copula functions, marginal distributions, covariate effects and data-driven variable selection.

The remainder of this manuscript is structured as follows: Section~\ref{sec:2} presents distributional copula regression for bivariate right-censored time-to-event as well as semi-competing risks responses and outlines our boosting algorithm. Section~3 documents our simulation studies and respective results. In Section~4 we analyse a high-dimensional ($p \gg n$) micro-array dataset related to patients suffering from ovarian cancer in which the time-to-event responses, time of tumour progression and time of death, follow a semi-competing risks data generating process. We model the joint survival function of the time of tumour progression and time of death as a function of  genomic as well as clinical information. Additionally, we illustrate the model-building process that involves selecting marginal distributions and the copula function. Lastly, a discussion is given in Section~5.

\section{Methods}\label{sec:2}

In this section, we briefly introduce right-censored and semi-competing risks time-to-event responses. Afterwards we outline our distributional copula regression framework for bivariate right-censored time-to-event responses and describe how to perform estimation by means of component-wise gradient boosting. 
\begin{figure}[!t]
     \centering
     \includegraphics[scale=0.65]{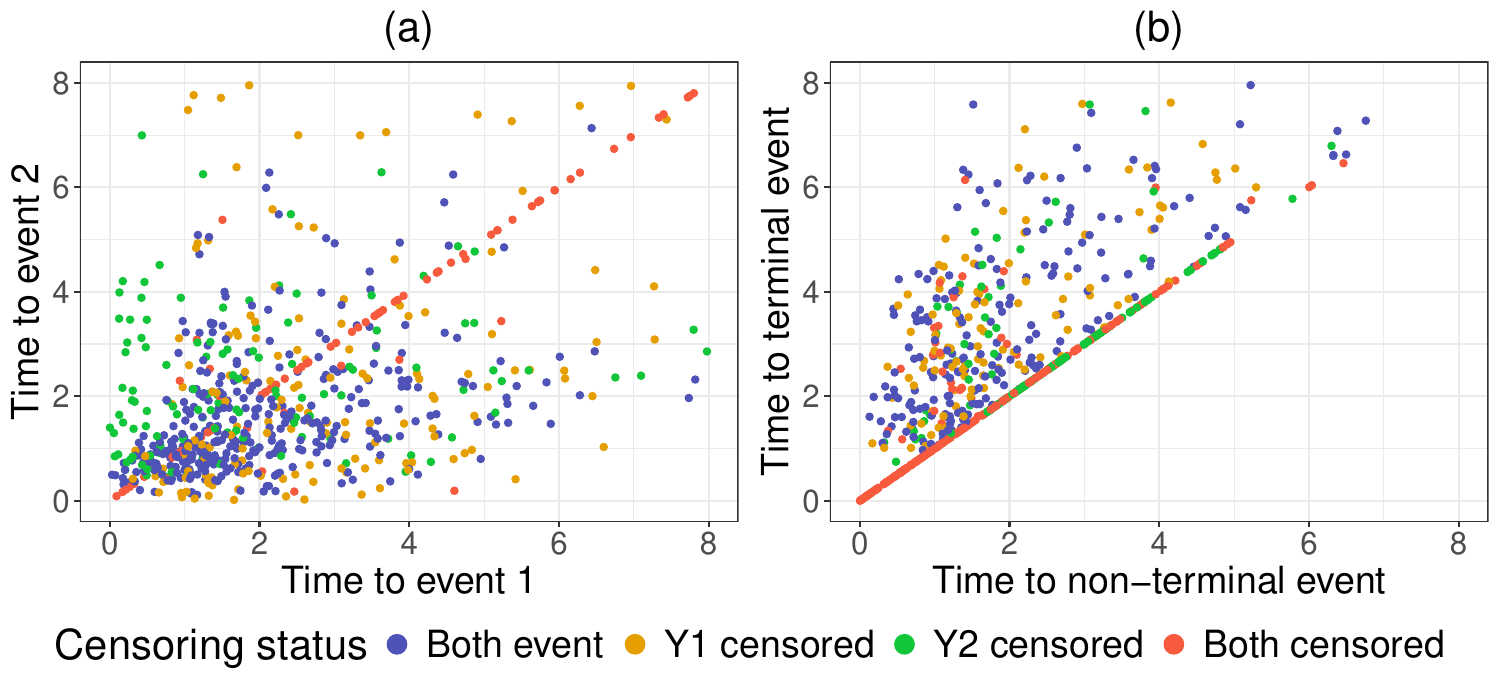}
     \caption{Synthetic bivariate time-to-event data with right-censoring~(a) and  semi-competing risks~(b).}
     \label{Figure:ExampleScatterplot}
 \end{figure}

\subsection{Right-censored time-to-event responses}\label{subsec:Semicompeting}

A univariate right-censored time-to-event response is comprised of $Y = \min\{T, \Tilde{T}\}$ and its censoring indicator $\delta = \mathds{1}\{T \leq \Tilde{T} \}$, where $T$ is the true event time and $\Tilde{T}_{}$ is an independent, random, uninformative censoring time. In addition, we assume that we have some covariate information $\xvec$ available. In what follows, we are concerned with bivariate right-censored time-to-event responses which consist of two univariate right-censored event times $\mY = (Y_{1}, Y_{2})^\top$ and their corresponding indicators $\deltavec=(\delta_{1}, \delta_{2})^\top$, and we write $(\mY,  \deltavec)$ for their pair. An example of simulated bivariate time-to-event data with right-censoring scheme is shown in Figure~\ref{Figure:ExampleScatterplot}(a). Throughout, we make the common assumption that the marginal censoring times remain independent of their respective true event times as well as from each other.

Moreover, we consider a special type of right-censored time-to-event outcome that naturally produces bivariate data known as ``semi-competing risks'' \citep[SCR;][]{FinJiaChap2001, Wang2003}. Semi-competing risks responses usually contain information about a non-terminal and a terminal event. The terminal event may censor the non-terminal one but it remains observable if the non-terminal event occurs first \citep{FinJiaChap2001}. In biomedical applications, the terminal event is typically death, whereas the notion of the non-terminal event time is usually a landmark event e.g.,\ time of disease progression. Using our notation, let the true non-terminal and terminal events be denoted by $T_1$ and $T_2$, respectively.  Semi-competing risks generate bivariate time-to-event data since one observes the first event $Y_{1} = \min\{T_1, T_2, \Tilde{T}\}$ with its corresponding censoring indicator $\delta_{1} = \mathds{1}\{T_1 \leq \min\{T_2, \Tilde{T}\} \}$. The second observed time-to-event response is then determined by $Y_{2} = \min\{T_2, \Tilde{T}\}$ as well as $\delta_{2} = \mathds{1}\{T_2 \leq  \Tilde{T} \}$ and we   again write $(\mY,  \deltavec)$ for their pair. Figure~\ref{Figure:ExampleScatterplot}(b) shows a scatterplot of simulated data with semi-competing risks responses.  

\subsection{Model structure}

To describe the entire conditional distribution of right censored time-to-event variables, we make use of a distributional copula regression approach based on generalized additive models for location, scale and shape \citep[GAMLSS;][]{GAMLSSPAPERORIGINAL}. Specifically, we follow \cite{Marra2019_CLBMODELS_JASA} and \cite{WeiWojSorr2024} and assume that the joint survival function $S(t_1,t_2;\varthetavec)=\dsP(T_1>t_1,T_2>t_2;\varthetavec)$ is given by 
\begin{align}\label{CopulaIsJointDistribution}
    S(t_{1},\ t_{2} ;\ \boldsymbol{\vartheta}) =  C[S_1(t_{1} ;  \boldsymbol{\vartheta}^{(1)} ), \ S_2(t_{2};  \boldsymbol{\vartheta}^{(2)}); \  \vartheta^{(c)} ],
\end{align}
where $C(\cdot, \cdot;\,\vartheta^{(c)}): [0,1]^2 \rightarrow [0,1]$ is a one-parameter bivariate copula function with association parameter $\vartheta^{(c)}\in\dsR$, and $S_1(t_{1} ;  \boldsymbol{\vartheta}^{(1)})=\dsP(T_1>t_1;\varthetavec^{(1)})$ and $S_2(t_{2} ;  \boldsymbol{\vartheta}^{(2)})=\dsP(T_2>t_2;\varthetavec^{(2)})$ are the possibly different univariate parametric marginal survival functions with respective distribution parameter vectors $\boldsymbol{\vartheta}^{(1)}\in\dsR^{K_1},\boldsymbol{\vartheta}^{(2)}\in\dsR^{K_2}$. Altogether, the bivariate joint survival function depends on the parameter vector $\varthetavec=(( \boldsymbol{\vartheta}^{(1)})^\top, (\boldsymbol{\vartheta}^{(2)})^\top, \vartheta^{(c)} )^\top \in \dsR^{K}$ with $K = K_1 + K_2 + 1$. 

\paragraph{Dependence measures} An advantage of resorting to copulas is the separation of specifying the marginal distributions and their respective dependence structure. This flexibility could help to uncover important aspects of the association between the marginal event times. In this context, relevant dependence measures are Kendall's $\tau$ rank correlation, upper and lower-tail dependence coefficients, and the cross-ratio function. The upper-tail dependence coefficient is defined as $\psi_U = \lim_{q \rightarrow 1} \dsP(t_2 > F_2^{-1}(q) | t_1 > F^{-1}(q))$, whereas the lower-tail dependence coefficient is given by $\psi_L= \lim_{q \rightarrow 0^{+}} \dsP(t_2 \leq F_2^{-1}(q) | t_1 \leq F^{-1}(q))$. For instance, the presence of lower-tail dependence would imply that the association between the margins is stronger at the end of the follow-up time (i.e.,\ when $S_1, S_2 \to 0$) and weaker close to the beginning of the study (i.e.,\ when $S_1, S_2 \to 1$), and vice versa for upper-tail dependence. The cross-ratio function is given by
\begin{equation*}
R_{\vartheta^{(c)}}(u_1,u_2) = \frac{ c(u_1,u_2 ;\,\vartheta^{(c)}) \ C(u_1, u_2;\,\vartheta^{(c)})}{{ C(u_1 \mid u_2;\vartheta^{(c)} )}\ { C(u_2\mid u_1;\vartheta^{(c)})}}
\end{equation*}
where 
$
c(\cdot,\cdot;\,\vartheta^{(c)})
$ 
denotes the copula density, $u_1=S_1(t_{1} ;  \boldsymbol{\vartheta}^{(1)})$, $u_2=S_2(t_{2} ;  \boldsymbol{\vartheta}^{(2)})$, and the terms $C(u_1 \mid u_2;\vartheta^{(c)} ) = \partial C(u_1, u_2;\,\vartheta^{(c)}) / \partial u_2$, and $C(u_2 \mid u_1;\vartheta^{(c)} ) = \partial C(u_1, u_2;\,\vartheta^{(c)}) / \partial u_1$ denote the conditional copula function given the margin $u_1$ or $u_2$, respectively. The cross-ratio function provides a measure of local dependence between the margins at $S_1{}, S_2$. Values of $R_{\vartheta^{(c)}} > 1$ indicate positive local dependence, whereas $0 < R_{\vartheta^{(c)}} < 1$ points toward negative local dependence. The special case of $R_{\vartheta^{(c)}} = 1$ corresponds to local independence \citep{EmuChen2018Book}.

\paragraph{Dependence structure}
We have implemented a wide range of copula functions such as the Gaussian, which is the most prominent example of elliptical copulas, as well as four Archimedean copulas (Frank, Gumbel, Clayton and Joe) with 0, 90\degree, 180\degree and 270\degree\, rotations of the latter three. Rotating the Clayton, Gumbel and Joe copulas results in changing the direction of the dependence structure to different parts of the quadrant. The three Archimedean copulas and their rotated versions, in contrast to the Gaussian and Frank copulas, do allow for tail dependence.

\paragraph{Marginal survival functions} Our implementation features the four most prominent parametric distributions for AFT models: Exponential, Weibull, log-logistic and log-normal. All of the implemented distributions depend on two scalar parameters. Tables~\ref{ImplementedAFTDistributions} and~\ref{ImplementedCopulas} summarize the currently implemented marginal distributions and copula functions, respectively.

\subsection{Predictor specifications}
 Each of the $K = K_1 + K_2 + 1$ parameters of the joint survival function, is modelled as a function of covariates using  structured additive predictors $\eta_{k}^{(\bullet)}$ of the form 
\begin{align}\label{DistRegApproachPredictor}
g_{k}^{(\bullet)}( \vartheta_{k}^{(\bullet)}) =\eta_{k}^{(\bullet)}  &= \beta_{0k}^{(\bullet)}  +  \sum_{r = 1}^{ P_{k}^{(\bullet)}} s_{ rk }^{(\bullet)} ( \xvec_{rk}), 
\quad  
\bullet\in\lbrace 1, 2, c\rbrace, \ 
k=1,\ldots,K_{\bullet}, \
\text{and } K_c=1,
\end{align}
where $\xvec_{rk}\subset\xvec$, and $g_k(\cdot)$ are link functions with corresponding inverse functions $h_k(\cdot)\equiv g_k^{-1}(\cdot)$,  guaranteeing that the individual parameters comply with their respective parameter space restrictions. The structured additive predictors $\eta_{k}^{(\bullet)}$ are composed of a parameter-specific intercept $ \beta_{0k}^{(\bullet)}$ and smooth functions of the covariates denoted by $s_{ rk }^{(\bullet)} ( \cdot )$. The latter can accommodate a wide range of functional forms, such as linear, non-linear and spatial effects. This is because each  $s_{ rk }^{(\bullet)} ( \cdot )$ is modelled through a linear combination of appropriate basis function expansions of the form
$$
s_{ rk }^{(\bullet)} (\xvec_{rk})=\sum_{l=1}^{L_{rk}^{(\bullet)}}\beta_{rk,l}^{(\bullet)}B_{rk,l}^{(\bullet)}(\xvec_{rk}),
$$ 
where $B_{rk,l}^{(\bullet)}(\xvec_{rk})$ are the basis functions evaluated at  $\xvec_{rk}$ and $\beta_{rk,l}^{(\bullet)}$ are the corresponding unknown regression coefficients which must be estimated, see \citet{Woo2017} for more details. 

The summation index $P^{(\bullet)}_k$ in Equation~\eqref{DistRegApproachPredictor} emphasizes that the subset of covariates assigned to each parameter do not need to be the same. In fact, it may be the case that no covariates have an effect on some parameters $\vartheta_{k}^{(\bullet)}$ of the joint survival function $S(\cdot,\cdot;\varthetavec)$. Thus, in general there may not be strong a-priori evidence of which subset of covariates (or if any at all) has an effect on the parameters of $S(\cdot,\cdot;\varthetavec)$. In order to tackle these model-building and variable-selection challenges in a data-driven manner, we resort to component-wise gradient-boosting or statistical boosting to estimate the model coefficients.

\subsection{Estimation via component-wise boosting}\label{BoostingSubsection}

Statistical boosting \citep{EvolutionOFBOOSTINGALGORITHMS} is based on a component-wise gradient boosting algorithm with regression-type {base-learners} \citep{FriedmanSOLO2000BOOSTING, BuehlmannHothornBOOSTING}. In our case, these base-learners correspond to the smooth components  $s_{rk}^{(\bullet)}(\xvec_{rk})$, $\bullet\in\lbrace 1, 2, c\rbrace$. A complete list of the currently implemented base-learners in the context of boosting can be found in \cite{Mayr2012GAMBOOSTLSS}. Let $\lbrace (\yvec_i, \deltavec_i,\xvec_i)\rbrace_{i=1}^n$ be the observed time-to-event data.

Then, estimation of the model coefficients is carried out by iteratively minimizing the empirical risk: $\omega_n=\frac{1}{n} \sum_{i=1}^{n} \omega(\boldsymbol{y}_i; \boldsymbol{\vartheta}_i)$, where $\boldsymbol{\vartheta}_i = (\boldsymbol{\vartheta}_i^{(1)}, \boldsymbol{\vartheta}_i^{(2)}, {\vartheta}_i^{(c)} ) \in \mathbb{R}^K$ is the distribution parameter vector for observation $i$, and  $\omega(\cdot;\cdot)$ represents the loss function of interest. In our case, the loss is equal to the negative log-likelihood of our model $\mathcal{L}=-\sum_{i=1}^n \ell_i$, where $\ell_i$ is the log-likelihood contribution. A single contribution to the log-likelihood is given by
% here we need some free line or the spacing gets weird

{\begin{footnotesize}
\begin{equation}\begin{aligned}\label{BIVAFTLOGLIK}
    \ell &= (1-\delta_{1})(1-\delta_{2}) \Big\{ \log(  C\lbrack S_1(y_{1}; \boldsymbol{\vartheta}_{}^{(1)} ),\  S_2( y_{2}; \boldsymbol{\vartheta}_{}^{(2)}) ;\ {\vartheta}_{}^{(c)} \rbrack ) \Big\} + \\
    %
    %
    %\\
    %
    %
    %
    &\phantom{=} (1-\delta_{1})\delta_{2}  \left\{  \log\left( \frac{ \partial C \lbrack S_1(y_{1}; \boldsymbol{\vartheta}_{}^{(1)}  ), \ S_2( y_{2}; \boldsymbol{\vartheta}_{}^{(2)} ); \  {\vartheta}_{}^{(c)} \rbrack }{ \partial S_2(y_{2}; \boldsymbol{\vartheta}_{}^{(2)} ) } \right) 
    \ + \ 
    \log( f_2( y_{2}; \boldsymbol{\vartheta}_{}^{(2)}) ) \right\}  +  \\
    %
    %
    %\\
    %
    %
    %
    &\phantom{=} \delta_{1}(1-\delta_{2})   \left\{  \log\left( \frac{ \partial C\lbrack S_1( y_{1} ; \boldsymbol{\vartheta}_{}^{(1)} ), \ S_2( y_{2}; \boldsymbol{\vartheta}_{}^{(2)} );\  {\vartheta}_{}^{(c)} \rbrack }{ \partial S_1( y_{1}; \boldsymbol{\vartheta}_{}^{(1)} ) } \right) 
    \ + \ 
    \log( f_1( y_{1}; \boldsymbol{\vartheta}_{}^{(1)}) ) \right\}  + \\
    %
    %
   %\\
    %
    %
    %
    &\phantom{=} \delta_{1}\delta_{2}  \Big\{  \log( c\lbrack S_1( y_{1}; \boldsymbol{\vartheta}_{}^{(1)} ), \ S_2( y_{2}; \boldsymbol{\vartheta}_{}^{(2)} );\  {\vartheta}_{}^{(c)} \rbrack ) + 
    \log( f_1( y_{1}; \boldsymbol{\vartheta}_{}^{(1)} ) ) + 
    \log( f_2( y_{2}; \boldsymbol{\vartheta}_{}^{(2)} ) ) \Big\},
\end{aligned}\end{equation} \end{footnotesize}}

% keep one line here
\noindent where the functions $f_1( y_{1}; \boldsymbol{\vartheta}_{}^{(1)} )$ and $f_2( y_{2}; \boldsymbol{\vartheta}_{}^{(2)})$ are the marginal probability density functions (PDFs). In each  iteration of the statistical boosting algorithm each of the pre-specified base-learners (components) of each distribution parameters is fitted individually to the negative gradient of the loss function w.r.t.\ to the additive predictors of the parameters. These quantities are also referred to as {pseudo-residuals}, and are given by $- \partial \omega(\boldsymbol{y}_i; \boldsymbol{\vartheta}_i) / \partial \eta^{(\bullet)}_{ k i }$. Based on a prediction criterion, only the best-performing base-learner or component out of all additive predictors is selected and a  ``weak'' update of the model is conducted \citep{Thomas2017NONCYLCIC}. The procedure is carried out for a pre-specified number of iterations denoted by $\texttt{m}_{\texttt{stop}}$. Conducting {early stopping}, i.e., using $\texttt{m}^{\texttt{opt}}_{\texttt{stop}} < \texttt{m}_{\texttt{stop}}$  iterations leads to some base-learners being effectively left out of the model. Hence statistical boosting conducts intrinsic, data-driven variable selection as well as shrinkage of the covariate effects. This implies that the number of fitting iterations $\texttt{m}_{\texttt{stop}}$ is the main tuning parameter. 

\paragraph{Implementation details} Our approach extends the boosting methodology presented in \cite{HansBoostDistReg} and \cite{BriKle2024} to bivariate right-censored time-to-event data. 
Estimation is carried out in a two-step fashion akin to \cite{Joe2005} described in detail in Algorithm~\ref{boostingAlgorithm}. In the first step, the coefficients of the sub-models of the margins are boosted separately, i.e.,\ an optimal number of fitting iterations is obtained for each marginal survival model  ($\texttt{m}^{\texttt{opt}(\bullet)}_{\texttt{stop}}, \bullet = 1, 2$). In the second step, we compute $\hat{S}_\bullet(y_{\bullet }; \boldsymbol{\hat\vartheta}_{}^{(\bullet)} )$, as well as $\hat{f}_\bullet(y_{\bullet }; \boldsymbol{\hat\vartheta}_{}^{(\bullet)})$ at the respective $\texttt{m}^{\texttt{opt} (\bullet)}_{\texttt{stop}}$ with $\bullet \in \{1, 2\}$ and plug them into the log-likelihood function shown in Equation~\eqref{BIVAFTLOGLIK}. The latter is then boosted as a function of $\vartheta_{}^{(c)}$. 

For data generated by SCR responses, we proceed similarly but boost only the margin of the terminal event $T_{2}$ and compute the fitted survival function and density at $\texttt{m}^{\texttt{opt} (2)}_{\texttt{stop}}$. In the second stage we plug the aforementioned functions into Equation~\eqref{BIVAFTLOGLIK} and boost it as a function of $\boldsymbol{\vartheta}^{(1)}$ and $\vartheta^{(c)}$. The algorithm has been integrated into the \textsf{R} package \texttt{gamboostLSS}. We denote our proposed approach described above by \texttt{SurvCopBoost}. Section~\ref{UsingSurvCopBoost} in the Supplementary Material provides an illustration on how to fit the proposed model class using the \texttt{SurvCopBoost} function implemented in \textsf{R}.

\section{Simulation study}\label{sec:3}

In this section, we conduct a number of experiments  to empirically evaluate the estimation accuracy, the predictive performance and the ability of our approach to conduct consistent variable selection. In our experiments, we consider $p_1=10$, $p_2 = 500$, $p_3 = 1000$, as well as different censoring regimes in two different scenarios: In Section~\ref{sec:SCR} we mimic semi-competing risks (SCR) data with censoring rates similar to those found in our application from Section~\ref{sec:4}. The simulations in Section~\ref{sec:BTE}  treat a bivariate time-to-event data (BTE) data generating process (DGP) with ``mild'' ($\approx30\%$) and ``heavy'' ($\approx70\%$) censoring rates in each margin. Before describing the two scenarios in detail, we state the following general settings that hold for both.

\subsection{General settings}

\paragraph{Data generation}
To build the bivariate response distributions we consider the Weibull and log-logistic distributions for the first and second margin, respectively. Bivariate samples from a copula are obtained using the package \texttt{VineCopula} \cite{VineCopulaPackage}. The copula and predictor choices are scenario-specific and discussed separately. The amount of censoring times also depends on the scenario but in both cases, censoring times are generated independently from univariate distributions. %The covariates are obtained from independent unit uniform distributions. 
The covariates are generated from a multivariate Gaussian distribution with Toeplitz covariance structure of the form $\Sigma_{ij} = \rho^{\mid i - j\mid}$ for $1 \leq i, j \leq p_q$, with $\rho = 0.5$ denoting the correlation between consecutive covariates $x_j$ and $x_{j+1}$. The range of each covariate is then transformed to the unit interval by means of the standard normal CDF. We generate 500 replicate training data sets of size  $n_{\texttt{train}} = 1000$ observations each and evaluate the performance on an additional test set of the same size denoted by $n_{\texttt{test}}$. Since we consider one-parameter copulas and the Weibull and log-logistic distributions come with two distribution parameters each, we have a total of $K=5$ distribution parameters throughout. Thus, it is worthwhile noting that the cases $p_2,p_3$ come with 2500 and 5000 potential covariates, such that both can be considered as \emph{high-dimensional} (i.e.\ $p>n$).

\paragraph{Performance  evaluations and benchmarking}  All performance evaluations are computed using the separate test set. The goodness-of-fit of \texttt{SurvCopBoost} is assessed using the negative log-likelihood (log-score) and compared against the respective scores of a competing model assuming the same but independent margins. For further comparison, we evaluate the performance for each margin separately (thus not evaluating the loss in ignoring potential dependence in the responses) in comparison with independent univariate Cox models, as they represent the most popular approach in survival analysis. To allow for a fair comparison, we used boosting to estimate the Cox models as well. Lastly, we include a penalised maximum likelihood approach implemented in the \texttt{GJRM} \cite{GJRMRPackage} \textsf{R} package. The respective criteria are the Integrated Brier Score (IBS), the Integrated Squared Error (ISE), the Integrated Absolute Error (IAE), the Concordance Index (C-Index), as well as true and false positive rates (TPR, FPR, respectively).  

\paragraph{Implementation details and tuning}
To carry out the weak learning mechanism of boosting, we need to set a sensible step-length $\texttt{s}_{\texttt{step}}$. Here, we follow  \cite{BriKle2024} and set $\texttt{s}_{\texttt{step}} = 0.1$ for all distribution parameters. However, in order to obtain similar step-lengths among the distribution parameters of the margins, we apply $L_2$-stabilisation to the parameter-specific gradients \citep{HofnerGAMBOOSTLSS}. We adopt the same step-length for the boosted independent Cox models. The stopping iteration $\texttt{m}_{\texttt{stop}}$ of \texttt{SurvCopBoost} and the independent Cox models is optimised by minimising the out-of-bag empirical risk on a further validation data set (different from the test data set) of size $n_{\texttt{mstop}} = 1000$ obtained from the same underlying distribution. We fitted all \texttt{SurvCopBoost} models in \textsf{R} using our implementations via the \texttt{gamboostLSS} package. The boosted Cox models are fitted using the implementation from the package \texttt{mboost} \citep{mboostPackage}. The code to reproduce all results is available on the following GitHub repository: \url{https://github.com/GuilleBriseno/BoostDistCopReg_Surv}.

\subsection{Semi-competing risks (SCR) responses}\label{sec:SCR}

\paragraph{Data generation} Motivated by the data analysed in Section~\ref{sec:4}, we generate bivariate time-to-event responses that follow the SCR mechanism described in Subsection~\ref{subsec:Semicompeting} with dependence structure based on a Gumbel copula. Based on the needs of the application, we assume linear predictors given by
\begin{align*}
    \log \vartheta_{i1}^{(1)} = \eta^{(1)}_{i1}\ &=  - 2x_{1i}
    , \\
    \log \vartheta_{i2}^{(1)} = \eta^{(1)}_{i2}\  &=  + 1x_{2i} + 1.5x_{4i} ,
     \\
     \log \vartheta_{i1}^{(2)} = \eta^{(2)}_{i1}\ &=  + 1 x_{1i} + 1.5 x_{2i}
    , \\
    \log \vartheta_{i2}^{(2)} = \eta^{(2)}_{i2}\  &=  +1 +0.75 x_{2i} + 0.75 x_{4i} ,
     \\
    \log( \vartheta^{(c)}_i - 1 )= \eta^{(c)}_{i} &=  3 -2 x_{2i} - 2 x_{4i},
\end{align*}
as well as censoring rates of $\approx40\%$ and $\approx47\%$ in each margin, respectively. The censoring times were sampled from a univariate uniform distribution on the interval $[0;\ 7]$. In this case only three out of the $p_q$, $q\in\lbrace 1,2,3\rbrace$ covariates have non-zero effects on the distribution parameters. Note that there is an overlap of the informative covariates between the different distribution parameters. The Gumbel copula is able to model upper-tail dependence, hence one would expect larger values of the marginal survival functions (earlier event times) to exhibit a stronger dependence compared to lower values (later event times). Averaging over the observations, the dependence between the margins in terms of Kendall's $\tau$ lies within $[0.187;\ 0.922]$, thus ranging between moderate and very strong positive dependence. 

 Besides benchmarking with independent models and univariate Cox models, we also compare two ways to estimate \texttt{SurvCopBoost}. The first estimates the margins separately using the two-step algorithm described in Algorithm~\ref{boostingAlgorithm} and is denoted as \texttt{SurvCopBoost} BTE (\textit{bivariate time-to-event}) estimation. The second estimates first the coefficients that correspond to the margins of the terminal event ($T_2$). Afterwards, the estimates $\hat{S}_2(\cdot)$ and $\hat{f}_2(\cdot)$ are plugged into Equation~\eqref{BIVAFTLOGLIK} and the remainder of the loss is boosted jointly. This procedure is denoted as \texttt{SurvCopBoost} SCR (\textit{semi-competing risks}) estimation. We remark that the estimation of the margin corresponding to the terminal event ($T_2$) is the same for both \texttt{SurvCopBoost} BTE and SCR estimation strategies. 

\paragraph{Results} Table~\ref{SimulationsSCRMetrics_TOEPLITZ} reports the performance metrics. Except the C-Index, all measures are oriented such that lower values indicate better performance. The reported scores are computed as the average of the 500 replicate test data sets. The results emphasize that our proposed \texttt{SurvCopBoost} leads to a better fit in terms of the log-score compared to ignoring the dependence structure and fitting independent models. This general observation holds true for both BTE and SCR estimation schemes. However, \texttt{SurvCopBoost} SCR appears to outperform the \texttt{SurvCopBoost} BTE estimation in terms of the log-score in low-dimensional settings $(p=10)$. In case of high-dimensional data  (i.e.,\ $p_2 = 500, p_3 = 1000$), the \texttt{SurvCopBoost} BTE strategy outperforms \texttt{SurvCopBoost} SCR in terms of the log-score. Univariate performance scores seem to favor \texttt{SurvCopBoost} BTE estimation compared to the \texttt{SurvCopBoost} SCR approach and also compared to fitting independent Cox models.

 Figure~\ref{SIM2_LINDGP_TOEP} displays the estimated linear effects of informative and non-informative covariates in the margin corresponding to the non-terminal event ($T_1$) as well as the dependence parameter $\vartheta^{(c)}$. In low-dimensional configurations ($p_1 = 10$), both \texttt{SurvCopBoost} BTE and \texttt{SurvCopBoost} SCR approaches perform similar in $\vartheta_{1}^{(1)}$ and $\vartheta^{(c)}$. The boxplots in Figure~\ref{SIM2_LINDGP_TOEP}, displaying coefficients resulting from \texttt{SurvCopBoost} BTE estimation, exhibit a small bias in the intercept as well as the informative covariates in the aforementioned parameters. For $p=500$ and $p=1000$ we see that the shrinkage effect on the parameter $\vartheta_{2}^{(1)}$ becomes stronger the more candidate covariates enter the model. The estimated coefficients of the terminal event are displayed in Figure~\ref{SIM2_LINDGP_M2ONLY_TOEP}. These boxplots show a similar pattern as those for the non-terminal event, i.e.,\ a stronger shrinkage of the covariate effects on $\vartheta_{2}^{(2)}$ as $p$ increases.

 Regarding the TPRs and FPRs, Table~\ref{Simulation2SelectionRatesSCR_TOEP} reveals that \texttt{SurvCopBoost} BTE estimation tends to select more non-informative covariates in the dependence parameter in low-dimensional configurations than \texttt{SurvCopBoost} SCR. On the other hand, for high-dimensional settings with $p_2 = 500$ or $p_3 = 1000$ potential covariates, \texttt{SurvCopBoost} BTE estimation also yields higher TPRs as compared to \texttt{SurvCopBoost} SCR. With the most notable differences in the selection rates being observed on the dependence parameter $\vartheta^{(c)}$. The implementation of \texttt{GJRM} could only be fitted using $p_1=10$ covariates. In that setting the corresponding FPRs were very high due to \texttt{GJRM}'s lack of variable selection mechanism. Other results obtained from \texttt{GJRM} are omitted.

\subsection{Bivariate right-censored time-to-event (BTE) responses}\label{sec:BTE}
\paragraph{Data generation} We consider two censoring regimes with average censoring rates of 30\% (``mild'') and 70\% (``heavy'') for both margins, respectively.  The bivariate observations are generated from a Clayton copula, which allows to model positive dependence as well as lower tail dependence between the margins. We consider two DGPs. The first DGP contains only linear effects of the covariates, whereas the second DGP consists of  non-linear effects. For these, the additive predictors are 
{\footnotesize
\begin{align*}
\text{Linear DGP:} & & \text{Non-linear DGP:}\\
   \log \vartheta_{1i}^{(1)}  &= \beta^{(1)}_{0,1} - 2x_{1i}  , & 
   \log \vartheta_{1i}^{(1)}  &= -1.8 \cos(4 x_{3i}),  \\
   \log \vartheta_{2i}^{(1)}  &=  + 1x_{2i} + 1.5x_{4i}, &
   \log \vartheta_{2i}^{(1)}  &= 0.02  -\sin(x_{1i}) + \exp(x_{1i} + 1)^2 + 3\cos(2 \pi x_{1i}) , \\ 
   \log \vartheta_{1i}^{(2)}  &=  \beta^{(2)}_{0,1} + 1 x_{1i} + 1.5 x_{2i}, & 
   \log \vartheta_{1i}^{(2)}  &= 2 \sin(4 x_{2i}), %
\\  
\log \vartheta_{2i}^{(2)}  &= \beta^{(2)}_{0,2} +0.75 x_{2i} + 0.75 x_{4i}  ,  & 
\log \vartheta_{2i}^{(2)}  &=  -0.979 \cos(2 x_{4i}) - 1.958 \tanh(x_{4i}),  \\ 
\log \vartheta_i^{(c)}  &= 3 -2 x_{2i} - 2 x_{4i} , &
\log \vartheta_i^{(c)}  &= -3.1 \cos(4 x_{3i}).%
\end{align*}}
Consequently, only three/four out of the $p_q$, $q\in\lbrace 1,2,3\rbrace$ covariates have  non-zero effects on the distribution parameters in the linear/non-linear DGPs, respectively. Furthermore, several of the few informative covariates have an effect on multiple distribution parameters which challenges estimation. For the linear DGP, the additive predictor of the dependence parameter $\vartheta^{(c)}$ covers Kendall's $\tau$ values  within $[0.159;\ 0.907]$, whereas for the non-linear DGP it ranges from $[0.022;\ 0.917]$. Thus covering from low to very strong positive dependence between $T_1$ and $T_2$ in both DGPs. In addition, the chosen intercepts $\beta^{(2)}_{0,1}$ paired with independent censoring times sampled from uniform distributions on $[0;\ 8.5]$ yield censoring rates of about  30\% and 70\% for the linear DGP. In the non-linear DGP, the mild censoring regime is obtained by using uniform distributions on $[0; 11]$, whereas the heavy censoring regime uses the interval $[0; 2.75]$ for sampling the censoring times.

\paragraph{Results for the linear DGP}  Table~\ref{SimulationsAllMetrics_TOEP} reports the log-scores. The difference in log-scores between \texttt{SurvCopBoost} and independent models starts to dissipate only in extreme cases with a very high number of potential covariates ($p_3 = 1000$) and heavy censoring in the margins (70\%), see column (2), $p_3 = 1000$. In line with these findings, \texttt{SurvCopBoost} also produces better univariate  scores  compared to the univariate Cox models. The estimated coefficients are shown in Figure~\ref{SIM1_LINDGP_TOEP}. Given a mild censoring rate (30\% in each margin) and a low number of potential covariates ($p_1 = 10$), \texttt{SurvCopBoost} recovers the effect of informative covariates quite well, although the shrinkage of effect estimates is stronger for the dependence parameter $\vartheta^{(c)}$. On one hand, increasing the number of potential covariates as well as increasing the censoring rate (70\%) has a negligible effect on the estimation of informative covariates on the distribution parameters $\vartheta_1^{(1)}, \vartheta_1^{(2)}$. On the other hand, the shrinkage of effect estimates  increases sharply in the parameter $\vartheta_2^{(2)}$ as well as the dependence parameter $\vartheta^{(c)}$. The parameter $\vartheta_1^{(2)}$ also exhibits considerable shrinkage of the effect estimates on high-dimensional settings and heavy censoring, although it is not as pronounced as on the two aforementioned parameters.

The TPRs and FPRs presented in the upper half of Table~\ref{Simulation1SelectionRates_TOEP} show that \texttt{SurvCopBoost} is able to accurately recover the effect of informative covariates across the studied configurations. It can be seen that the degree of shrinkage and regularization depends more on the censoring rate than on the number of potential covariates present in the data, e.g.,\ compare the TPR in columns (1) against (2) for $p_3=1000$ in Table~\ref{Simulation1SelectionRates_TOEP}. Similar to Section~\ref{sec:SCR}, the implementation of \texttt{GJRM} could only be fitted in configurations with $p_1=10$ covariates. The respective FPRs exhibited the same pattern as in Section~\ref{sec:SCR}. Once again, further results obtained using \texttt{GJRM} are omitted. 

\paragraph{Results for the non-linear DGP} Similar to the linear DGP, \texttt{SurvCopBoost} outperforms the independent models in terms of the log-score in almost all considered configurations. Under a high censoring rate (70\%) combined with a high number of potential covariates ($p_2 = 500, p_3 = 1000$ in Table~\ref{SimulationsAllMetrics_TOEP}) the performance of both models is similar. This behaviour can be also observed in some of the univariate scores such as the IBS and C-Index, where those produced by Cox models are slightly better than those from \texttt{SurvCopBoost}. The estimated non-linear effects of the informative covariates shown in Figure~\ref{SIM1_NLDGP_TOEP} indicate that the censoring rates and the increasing number of potential covariates have a negligible effect on the accuracy of the estimated effects on the parameters of the marginal survival functions. However, increasing amount of censoring and noise variables induces a stronger shrinkage of the estimated effects and thus a larger bias in the dependence parameter $\vartheta^{(c)}$. For example, the green curves in Figure~\ref{SIM1_NLDGP_TOEP} corresponding to the row showing 70\% censoring exhibit a flatter shape of the estimated non-linear effect compared to the row depicting 30\% censoring. 

The selection rates corresponding to the non-linear DGP are shown in the lower half of Table~\ref{Simulation1SelectionRates_TOEP}. As already established in the linear DGP, \texttt{SurvCopBoost} identifies the informative covariates in all parameters of the joint survival function regardless of the number of candidate covariates in the model in a mild censoring regime (30\% censoring). The FPR in low-dimensional settings are rather high for both \texttt{SurvCopBoost} and independent Cox models, but they rapidly shrink towards zero once a large number of candidate covariates enter the model. However, the TPR from the Cox models is considerably lower than those of \texttt{SurvCopBoost}. Results from  \texttt{GJRM}  are once again omitted and the FPR behave in the same way as described in the results of Section~\ref{sec:SCR}.

\subsection{Summary of the simulation results} 

Overall, \texttt{\texttt{SurvCopBoost}} demonstrated satisfactory results for both SCR and BTE data. It is able to effectively detect and recover all true effects across the distribution parameters of the bivariate distribution. However, a larger bias in the estimation of the dependence parameter under heavy-censoring has to be acknowledged. This is likely because the copula dependence parameter $\vartheta^{(c)}$ shows stronger shrinkage of informative effects compared to other parameters. The strength of induced shrinkage and regularization is also  influenced by the censoring rate and the number of candidate variables. This phenomenon may be attributed to the greedy nature of the algorithm, since a reduction of the loss from including a covariate with a small coefficient in the dependence parameter might not be large enough compared to updating a coefficient in any other parameter corresponding to the margins or even the intercept of $\vartheta^{(c)}$, i.e.\ constant dependence. 

In high-dimensional SCR configurations, such as the one analysed in Section~\ref{sec:4}, our proposed two-step estimation approach (\texttt{SurvCopBoost} BTE) performs well at identifying informative covariates as well as modeling the underlying bivariate distribution. Overall, evaluating the predictive behaviour via probabilistic scores  highlights the added value of the bivariate \texttt{SurvCopBoost} model compared to using boosting for independent AFT models or more traditional Cox models for bivariate time-to-event data. Compared to the penalised maximum likelihood approach of \texttt{GJRM}, the proposed \texttt{SurvCopBoost} allows not only for a more streamlined model-building process by selecting the most informative variables in a data-driven manner, but also for feasible estimation in high-dimensional ($p \gg n$) settings.

\section{Analysis of high-dimensional ovarian cancer data with semi-competing risks responses}\label{sec:4}

In this section we showcase the ability of the proposed \texttt{SurvCopBoost} to conduct data-driven variable selection in a challenging high-dimensional data structure with semi-competing risks responses. The data analysis is related to ovarian cancer, a leading cause of cancer death in women \citep{SieMilKim2020} and the second global cause of death from gynecologic cancers \citep{Bai2020}. We are concerned with estimating the joint survival function of the time to tumour progression, i.e.,\ a landmark event of the disease, and the time of death. Using \texttt{SurvCopBoost}, the parameters of the joint survival function are modelled as functions of informative covariates selected in a data-driven fashion from a high-dimensional covariate vector. The data were obtained from the \textsf{R} Bioconductor package \texttt{curatedOvarianData} \citep{Ganzfried2013}. Next, we describe the data extraction process, configurations used for the \texttt{SurvCopBoost} model, as well as the results of our analysis. 

\paragraph{Data structure} The data is comprised of four annotated studies (GSE17260, GSE30161, GSE9891, and TCGA) included in the \texttt{curatedOvarianData} \citep{Ganzfried2013} package. The studies were extracted according to the \texttt{patientselection.config} file, see the package's vignette for more details. Our extracted sample consists of a total of $n = 822$ patients. Following a semi-competing risks data generating process, the responses are given by each patient's time of tumour progression (non-terminal event, $T_1$) and their respective time of death or survival time (terminal event, $T_2$) after surgery. The time scale of the responses is given in days. The median time-to-event times are 570 and 1353 days, respectively. The censoring rate for tumour progression is $\approx 40\%$, whereas in $\approx 48\%$ of patients the terminal event was not observed. These censoring rates are similar to those considered in our simulations under an SCR DGP. We consider all the covariate information that is commonly available across the aforementioned studies. Information from the common covariates may be split into two types: genomic and clinical. The regressors containing genomic information are a total of $11{,}761$ gene expressions. Following \cite{Ganzfried2013} as well as \cite{EmuNak2018}, the independent variables with clinical information are the tumour stage according to the FIGO staging system (I-IV, dummy encoding) and the residual tumour size at surgery encoded as a dummy variable as well (0= under 1cm, 1 = over 1cm). This yields a total of $p = 11{,}763$ covariates, which corresponds to a high-dimensional setting. In fact, fitting a statistical model to such a data structure ($p\gg n$) is infeasible with standard techniques. A previous analysis conducted on similar data by \cite{EmuNak2018} carried out variable selection based on univariate hypothesis testing prior to model fitting \citep{JenKuoSto2002}. Their approach selected 158 gene expressions associated with the non-terminal event ($T_1$), and 128 genes for time of death ($T_2$) out of the same set of potential covariates we examine here. Afterwards a composite covariate \citep{Tukey1993} is taken as a summary of the selected ``most significant'' variables. In our case \texttt{SurvCopBoost} allows the entire covariate vector to enter the model directly. 
\begin{figure}[!t]
    \centering
    \includegraphics[scale=0.54]{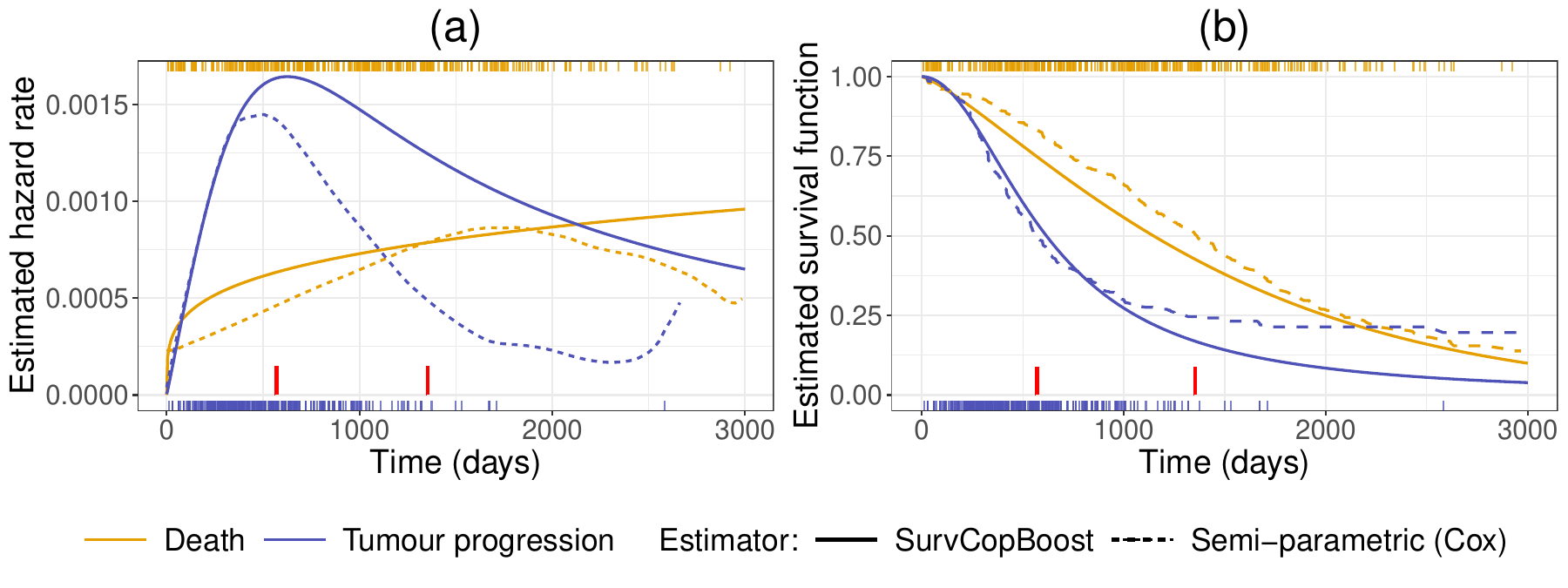}
    \caption{Estimated baseline hazard rate (a) and survival function (b) for time of tumour progression and time of death. Solid lines are estimates from \texttt{SurvCopBoost}, whereas dashed lines denote semi-parametric estimates corresponding to independent univariate Cox models. Thick red vertical lines highlight the median time of tumour progression (570 days) and median time of death (1353 days).}
    \label{Fig:EstimatedBaselineFunctions}
\end{figure}

\paragraph{Model configuration and tuning} We split our extracted sample into three partitions. The training data ($n_{ \texttt{train}}=577$), the validation data for tuning the number of fitting iterations ($n_{ \texttt{mstop}}=128$) and the data determining the optimal marginal distributions and copula function by means of the out-of-sample log-score ($n_{ \texttt{test} } = 117$), respectively.  The log-logistic, log-normal and Weibull distributions are considered as candidates for the margins of each of the event-times. For the dependence structure we fit 14 different implemented copula functions to the best-fitting marginal distributions. The copulas are the Gaussian, Frank, Clayton, Gumbel and Joe, as well as 90\degree, 180\degree and 270\degree\, rotations of the latter three. In total, the joint survival function consists of five parameters. 
The following additive predictor configuration is used for all parameters of the joint survival function:
\begin{equation}
    \eta_k^{(\bullet)} = \beta_{0k}^{(\bullet)}  +  \sum_{r = 1}^{ P_{k}^{(\bullet)}}  x_{rk} \beta_{ rk }^{(\bullet)} , 
\quad  
\bullet\in\lbrace 1, 2, c\rbrace, \ 
k=1,\ldots,K_{\bullet}, \
\text{and } K_c=1, 
\end{equation}
where $x_{rk}$ denotes one of the $p = 11{,}763$ covariates in the data. Hence, all covariates are modelled as linear functions. To the best of our knowledge, this is the first instance where the entire covariate vector is considered for modelling of this data. We determine the best-fitting marginal distributions and copula function by means of the out-of-sample log-score.  

Due to the relatively small sample size used for estimating the model coefficients ($n_{\texttt{train}}=577$), we set the step-length to $\texttt{s}_{\texttt{step}} = 0.005$. This configuration will lead to a larger number of optimal iterations, but it will keep the boosting algorithm stable throughout the fitting process. We apply $L_2$ stabilisation to the negative gradients of the loss and fit \texttt{SurvCopBoost} as stated before. Lastly, we fit independent univariate Cox models using boosting to each of the time-to-event responses for comparison.

\paragraph{Results} \noindent The best-fitting distribution for time to tumour progression is the log-logistic distribution, whereas for time of death it is the Weibull distribution. This result points to the difference in statistical behaviour between the time of tumour progression and the survival time. Figure~\ref{Fig:EstimatedBaselineFunctions}(a) shows the estimated baseline hazard rates as well as baseline survival functions 
 in (b). An important aspect is the mode of the hazard of time to tumour progression which can be seen to occur within the first 1000 days. This indicates a higher risk of tumour progression earlier after surgery compared to later in time. In contrast, the estimated baseline hazard of time to death has a monotonic increasing shape. The estimated baseline survival functions reveal the lower median time-to-event for the non-terminal event compared to death. Thus the drop in progression-free survival is much sharper compared to the terminal event. The estimated semi-parametric baseline hazard and functions that correspond to the Cox model follow those estimated by \texttt{SurvCopBoost} when there is a high prevalence of observations. The semi-parametric estimators show lower hazards in regions without observations, however this behaviour is expected in estimators of this type, see the rugs in Figure~\ref{Fig:EstimatedBaselineFunctions}(a) and (b). A similar phenomenon can be seen in the estimated semi-parametric baseline survival functions (dashed lines) in Figure~\ref{Fig:EstimatedBaselineFunctions}(b).

A total of $95$ covariates for the model of time to tumour progression (non-terminal event) is selected, see Table~\ref{Table:SelectedCovariates}. More specifically, it selects $73$ variables for the parameter $\vartheta^{(1)}_1$ and 24 variables for $\vartheta^{(1)}_2$ with only two genomic variables overlapping. The binary variable \texttt{residual tumour size} was the only clinical covariate selected for the sub-model $\vartheta^{(1)}_{1}$ of tumour progression. Our proposed \texttt{SurvCopBoost} and the significance-testing-based variable selection approach from \cite{EmuNak2018} have an overlap of 22 gene expressions. Out of these 22 overlapped variables, \texttt{SurvCopBoost} selects six of the top ten ``most significant'' expressions. 

Previous analyses and meta-analyses have shown the expression of gene $CXCL12$ (encoding a chemokine related to immune response) to be associated with survival \citep{Popple2012, Ganzfried2013, EmuNak2017}. Albeit these studies focused exclusively on this particular gene  while ignoring others. In this case \texttt{SurvCopBoost} selected $CXCL12$ only for the parameter $\vartheta^{(1)}_{1}$ of time to tumour progression's distribution. The association of this gene expression with the non-terminal event is also confirmed by \cite{EmuNak2018}. Other selected genes include members of the $TIMP$ family ($TIMP2$), which have functions associated with cell proliferation and survival \cite{BouJen2011}. The expression $PTPN4$, which has been found to perform an essential role in most phenotypes of tumour cells \citep{Tang2022}, was selected in both parameters $\vartheta^{(1)}_{1}$ and $\vartheta^{(1)}_{2}$ of the non-terminal event's distribution. Another gene selected in the aforementioned parameters was $FAT2$, which according to \cite{Wang2022} shows promise to be a predictor for responsiveness to immunotherapy and prognosis in uterine corpora malignant tumours. The gene $HIST1H4E$, selected for $\vartheta^{(1)}_{2}$, has been found to play a role in the production of CD8$^{+}$ regulatory T-cells or pathogen-combating cells \citep{Wu2016}.
\begin{table}[t!]
\centering
\caption{Out-of-sample log-scores of candidate marginal distributions and copula functions. Best-fitting values highlighted with bold numbers.}
\begin{tabular}{r rlcc r}
  \toprule
   \multicolumn{6}{c}{Selection of marginal distributions} \\
   \\
&& Distribution & tumour progression & Death &\\ 
&&  & (Non-terminal event, $T_1$) & (Terminal event, $T_2$) &\\ 
  \midrule
   \rowcolor{Gainsboro!60}
&& Weibull & 571.48 & \textbf{485.73}& \\ 
&& Log-logistic & \textbf{550.45} & 485.98& \\ 
 \rowcolor{Gainsboro!60}
&& Log-normal & 555.69 & 506.22 &\\ 
\end{tabular}\\
\begin{tabular}{rlr ccc clr}
  \midrule
    \midrule
  \multicolumn{9}{c}{Selection of copula function} \\
  \\
   & Copula & log-score &&&& &  Copula & log-score\\ 
  \midrule
   \rowcolor{Gainsboro!60}
1 & Independence & 1036.18          &&&&  9 & Clayton 90\degree & 1037.78\\ 
  2 & Gaussian & 1016.71            &&&& 10 & Gumbel 90\degree & 1037.42\\ 
   \rowcolor{Gainsboro!60}
  3 & Clayton & 1022.92             &&&& 11 & Joe 90\degree & 1037.16 \\
  4 & Clayton 180\degree & 1017.07  &&&& 12 & Clayton 270\degree & 1037.33 \\ 
   \rowcolor{Gainsboro!60}
  5 & Gumbel & \textbf{1014.70}     &&&& 13 & Gumbel 270\degree & 1037.78\\ 
  6 & Gumbel 180\degree & 1019.01   &&&& 14 & Joe 270\degree & 1037.82\\ 
   \rowcolor{Gainsboro!60}
  7 & Joe & 1017.39                 &&&& 15 & Frank & 1020.41 \\ 
  8 & Joe 180\degree & 1023.53      &&&& & & \\ 
  \midrule
    \multicolumn{9}{l}{log-scores computed using $n_{\texttt{test}}=117$ observations.} \\
  \bottomrule
\end{tabular}\label{Table:SelectionMarginCopula}
\end{table}

{\footnotesize
\begin{table}[t!]
\centering
\caption{Number of selected covariates and optimal fitting iterations of the parameters of the joint survival function using \texttt{SurvCopBoost} as well as boosted univariate independent Cox models. The symbols $\lambda_1$ and $\lambda_2$ denote the hazard rate corresponding to each Cox model.}
\resizebox{\columnwidth}{!}{%
\begin{tabular}{l cc c cc c c c cc}
  \toprule
& \multicolumn{2}{c}{Time of tumour progression ($T_1$)} && \multicolumn{2}{c}{Time of death ($T_2$)} && Dependence && Cox $T_1$ & Cox $T_2$ \\
& \multicolumn{2}{c}{Log-logistic distribution} && \multicolumn{2}{c}{Weibull distribution}  && Gumbel copula & &&\\
    \\
& $\vartheta^{(1)}_1$ & $\vartheta^{(1)}_2$ && $\vartheta^{(2)}_1$ & $\vartheta^{(2)}_2$ && $\vartheta^{(c)}$ && $\lambda_1$ & $\lambda_2$\\ 
  \midrule
Selected & \multirow{2}{*}{73} & \multirow{2}{*}{24} && \multirow{2}{*}{26} & \multirow{2}{*}{8} && \multirow{2}{*}{1} && \multirow{2}{*}{69} & \multirow{2}{*}{115} \\
covariates &  &  &&  &  &&  \\
 \rowcolor{Gainsboro!60}
%&&&&&&&\\
$\texttt{m}_{\texttt{stop}}^{\texttt{opt}}$ & 1740 & 2165 && 824 & 1313 && 18 && 2594 & 7487  \\
%&&&&&&&\\
Link  & $\ln(\cdot)$ & $\ln(\cdot)$ && $\ln(\cdot)$ & $\ln(\cdot)$ && $\ln(\cdot - 1)$ & & $\ln(\cdot)$ & $\ln(\cdot)$ \\
  \midrule
 \multicolumn{8}{l}{$L_2$ stabilisation, $\texttt{s}_{\texttt{step}}=0.005$, $n_{\texttt{train}}=577$, and $n_\texttt{mstop}=128$.}\\
   \bottomrule
\end{tabular}
}\label{Table:SelectedCovariates}
\end{table}}

For the distribution of the survival time a total of $34$ covariates were selected. As shown in Table~\ref{Table:SelectedCovariates}, out of the informative variables for the terminal event, 26 were selected for $\vartheta^{(2)}_1$ and eight for $\vartheta^{(2)}_2$, respectively. In this case there was no overlap in the selected covariates across the parameters. As previously mentioned, the univariate significance-testing-based variable selection approach used in  \cite{EmuNak2018} identified a total of 128 genes with time of death, which is a slightly sparser model compared to that of the non-terminal event (tumour progression). In our case we observe a similar pattern of a sparser model for the time of death. \texttt{SurvCopBoost} has twelve gene expressions in common with the approach from \cite{EmuNak2018} and features once again six variables of the top ten ``most significant'' ones. An important expression that was selected out of the most significant ones from \cite{EmuNak2018} is $TEAD1$. It has been found that the $TEAD$ genetic family is abnormally expressed in patients with Ovarian Serous Carcinoma \citep{Ren2021}, which is the most common type of ovarian cancer \citep{OCRA2021}. The selected expression of gene $YWHAB$ is associated with advanced stages of ovarian cancer as well as poor patient prognosis \cite{Li2021}. Our proposed \texttt{SurvCopBoost} selects $VSIG4$ into the sub-model $\vartheta^{(2)}_2$. It has been found that $VSIG4$ shows over-expression in ovarian cancers compared with benign tumours and could be a potential target for therapy \cite{Byun872}.

\begin{figure}[!t]
    \centering
    \includegraphics[scale=0.47]{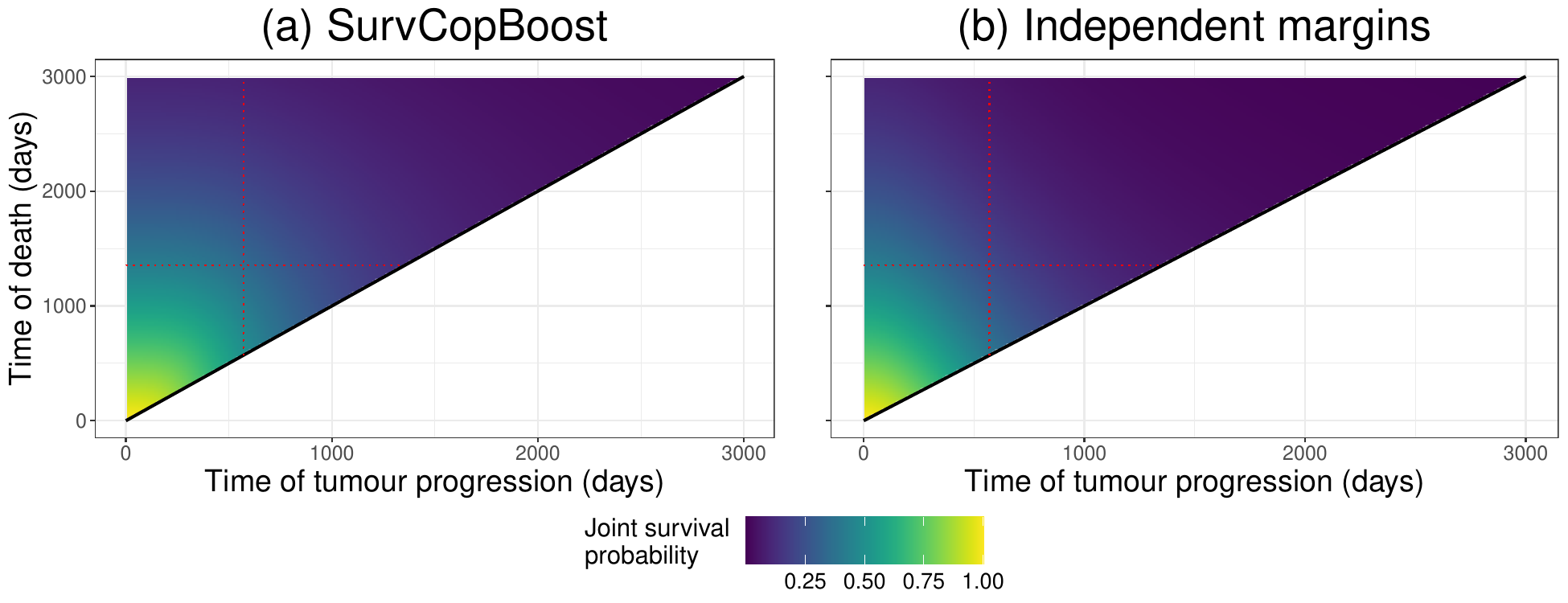}
    \includegraphics[scale=0.42]{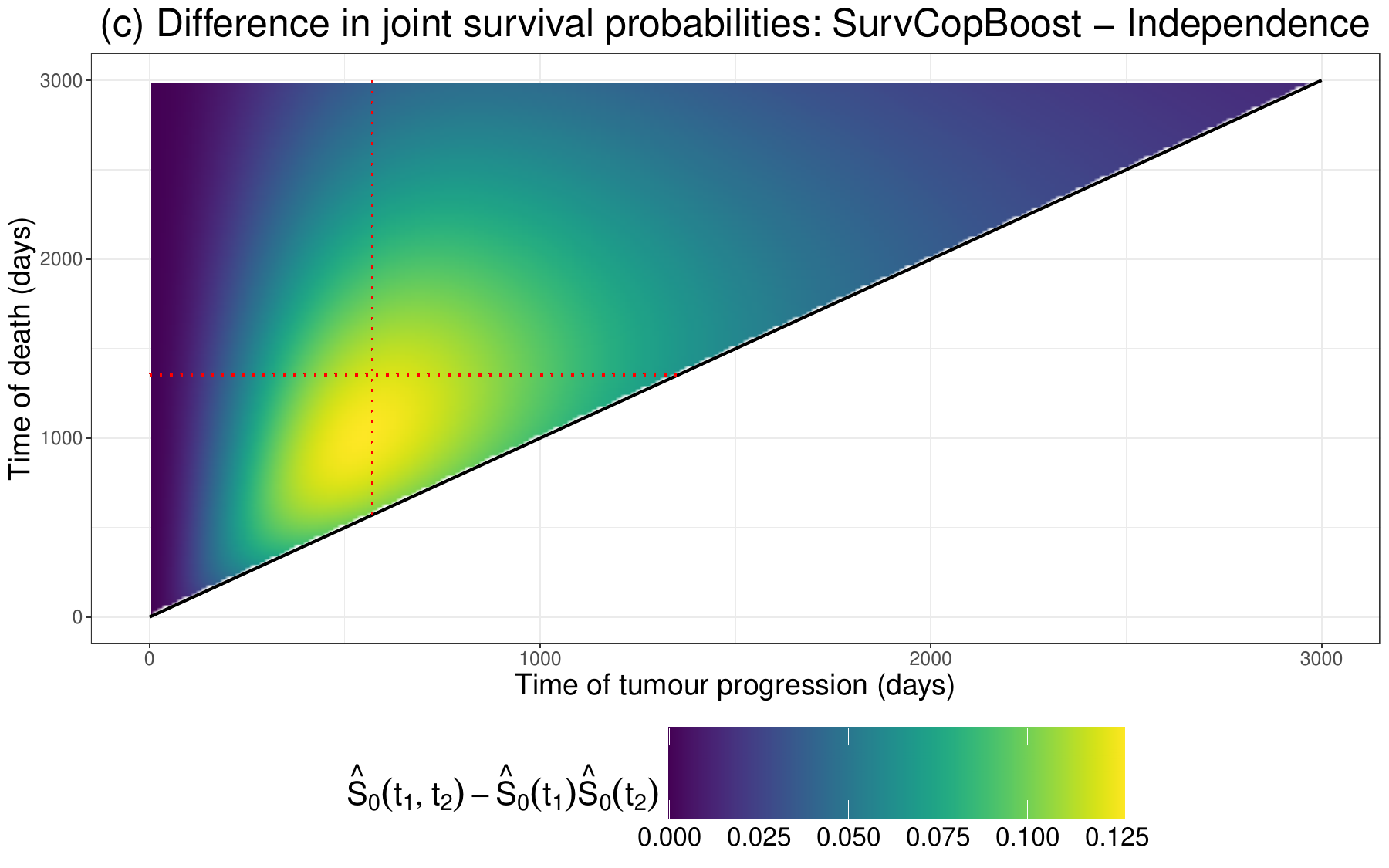}
    \caption{Estimated baseline joint survival probability of time of tumour progression and time of death in days with Gumbel copula using \texttt{SurvCopBoost} (a) as well as independent Log-logistic and Weibull margins (b). Difference between the baseline joint survival functions obtained using \texttt{SurvCopBoost} and independent margins, i.e.\ $\hat{S}_0(t_1, t_2; \hat{\varthetavec}) - \hat{S}_0(t_1; \hat{\varthetavec}^{(1)})\hat{S}_0(t_2; \hat{\varthetavec}^{(2)})$ (c). Red dotted lines indicate the median event-times: 570 days for tumour progression and 1353 days for death, respectively. Only the upper-wedge is defined for SCR data.}
    \label{Fig:EstimatedJointBaselineWedge}
\end{figure}

The Gumbel copula is selected as best-fitting dependence function, see Table~\ref{Table:SelectionMarginCopula}. This table furthermore reveals that the data strongly rejects copulas that support dependence for large values of time such as the Clayton, Gumbel 180\degree, or Joe 180\degree. Copulas that support negative dependence are strongly rejected as well. This can be seen in the worse predictive performance compared to that of a model with independent margins, see the log-score corresponding to 90\degree\ and 270\degree\  rotations. Figure~\ref{Fig:EstimatedJointBaselineWedge}(a) depicts the estimated baseline joint survival function according to the Gumbel copula model with log-logistic distributed time to tumour progression and Weibull distributed time of death. It can be seen that the joint survival is rather high for the first 100 days after surgery. A decrease in joint survival can be seen after 1000 days. The joint survival function assuming independent margins is shown in Figure~\ref{Fig:EstimatedJointBaselineWedge}(b). It can be seen that for regions close to the median event times the joint survival function assuming independence exhibits lower joint survival probabilities, compared to that of \texttt{SurvCopBoost}. The difference between the estimated joint survival functions, i.e.\ $\hat{S}_0(t_1, t_2; \hat{\varthetavec}) - \hat{S}_0(t_1; \hat{\varthetavec}^{(1)})\hat{S}_0(t_2; \hat{\varthetavec}^{(2)})$, is depicted in Figure~\ref{Fig:EstimatedJointBaselineWedge}(c). This shows that the joint survival probability of tumour progression and death is underestimated when both event times are modelled independently, with the biggest discrepancy between the estimates being observed close to the median event times, see the bright yellow spot around the intersection of the red dotted lines in Figure~\ref{Fig:EstimatedJointBaselineWedge}(c). 

Only one gene expression ($SLC16A10$) is selected for the model of $\vartheta^{(c)}$. This covariate was neither selected in the model of time to tumour progression nor in the one of time to death. Members of the $SLC16A$ gene family are important for cell metabolism \citep{Halestrap2004} and are known to play a crucial role in the process of tumourigenesis, i.e.,\ the formation of cancer as well as tumour progression \citep{Yu2020}. This particular variant was not selected by the significance-testing-based heuristic employed in \cite{EmuNak2018}. \texttt{SurvCopBoost} allows us to compute dependence measures in order to gain additional insights of the relationship between the margins.  The estimated baseline dependence between the margins expressed as Kendall's $\tau$ is $\hat{\tau} = 0.5$ and taking $SLC16A$ into consideration yields values of $\hat{\tau} \in [0.496;\ 0.510]$, indicating a moderate dependence between time to tumour progression and survival time. This result aligns with the estimated dependence previously found by \cite{EmuNak2017} and \cite{EmuNak2018}. Additionally, the Gumbel copula supports upper-tail dependence, thus meaning that the margins are dependent for extremely high values of their respective survival functions, i.e.,\ at very early times. This result is clinically reasonable, since patients that unfortunately suffer from tumour progression early after surgery typically also have a poorer prognosis of overall survival.

The range of the estimated upper-tail dependence coefficients in the data is $\hat{\psi}_U \in [0.582;\ 0.595]$. This shows that the margins are moderately dependent at extremely early times. In fact, the upper-tail dependence is higher than the dependence quantified by the estimated Kendall's $\tau$. Lastly, the values of the estimated cross-ratio function $\hat{R}_{\vartheta^{(c)}}$ show that the local dependence between the margins is always positive and becomes very high for some observations. The range of the estimated function is within $\hat{R}_{\vartheta^{(c)}} \in [1.230;\ 540.092]$ and has a median of $\text{med}(\hat{R}_{\vartheta^{(c)}}) = 2.333$.

\section{Discussion }\label{sec:5}

We have introduced \texttt{SurvCopBoost}, which is a distributional copula regression approach for bivariate time-to-event data under right-censoring and for semi-competing risks. Estimation in \texttt{SurvCopBoost} is carried out via statistical boosting \citep{BuehlmannHothornBOOSTING}. This enables data-driven variable selection, a feature that considerably simplifies the complex model building process. Our simulation studies show that \texttt{SurvCopBoost} outperforms other approaches (independent univariate boosted Cox and AFT, as well as bivariate copula time-to-event using penalised maximum likelihood) in terms of probabilistic forecast and exhibits similar performance to its competitors in terms of univariate metrics. \texttt{SurvCopBoost} also performs satisfactory in terms of variable selection by being able to identify informative covariates, as reflected TPRs and FPRs. All of these qualities were observed under different censoring regimes and growing number of noise variables in the model.

We analysed a high-dimensional data structure extracted from the \textsf{R} Bioconductor package \texttt{curatedOvarianData} \citep{Ganzfried2013} with time-to-event responses following a semi-competing risks data generating process. \texttt{SurvCopBoost} selected a subset of 129 informative covariates for the distributions of the marginal event times out of a potential $11{,}763$ variables. Therefore, \texttt{SurvCopBoost} demonstrates the benefit of conducting data-driven variable selection by analysing jointly the \textit{entire} covariate vector instead of relying on heuristics, for example hypothesis testing performed on univariate regression models. We believe that our application presented in Section~\ref{sec:4} demonstrates the advantages of using \texttt{SurvCopBoost} for analysing challenging data structures in a time-to-event analysis context. 

Currently \texttt{SurvCopBoost} implements three parametric distributions: Weibull, log-logistic and log-normal. The implementation of ``umbrella'' distributions, such as the generalised gamma \citep{CoxChuSch2007} or generalised $F$ distributions \citep{Cox2008}, which contain the already implemented ones as special cases, could be an option to further extend the flexibility of \texttt{SurvCopBoost}. A potential caveat of the current implementation of \texttt{SurvCopBoost} is the distributional assumption of a specific family for the marginal event times. Identifying a suitable distribution might be challenging in some cases. A pragmatic solution could be to implement Cox-type margins \citep{DerVan2024} or fully non-parametric margins \citep{Akri2004}. However, we consider link-based or ``generalised time-to-event models'' \citep{XinPaw2018, Marra2019_CLBMODELS_JASA} to be a more appropriate approach since those models are based on semi-parametric regression techniques.

We are currently exploring the inclusion of cure fractions, i.e.,\ cure models \citep{OthBar2012, Peng2021}, to account for observations that do not experience the event of interest, or in other words, their survival function does not reach zero. For example, this can be the case in semi-competing risk data where there are individuals that will not experience the landmark or non-terminal event. Combining statistical boosting and cure models can be very beneficial, since it is likely that some covariates will have an effect on the cure fraction and not on the survival function or vice versa. Therefore a purely data-driven variable selection mechanism could simplify the model building process. Other areas of active research are the censoring scheme and mechanism or their underlying assumptions thereof. We are interested in adapting a more general censoring scheme, which would allow to model data that features not only right, but also left and interval-censored observations, see e.g.,\ \cite{SunDing2019} or \cite{PetEle2022}. Regarding the censoring mechanism, the validity of the independent, as well as non-informative censoring in the marginal responses can be put up to debate / openly challenged or questioned. Allowing for dependent censoring in the marginal responses would require us to model the dependence structure between the marginal censoring and event times, see e.g., \cite{CzaKei2021}. Informative censoring could be addressed by adapting the approach of \cite{InformativeCensoringGJRM} to the framework of \texttt{SurvCopBoost}. These developments would result in a more complex model structure but will ultimately be beneficial for practical data analysis.

The boosting algorithm underlying \texttt{SurvCopBoost} is prone to some shortcomings. One of these aspects is the rather high FPRs, i.e.\ including non-informative explanatory variables in the model, in particular in low-dimensional settings. De-selection of non-informative covariates as proposed by \cite{StroemerDESELECTION} for statistical boosting could be adopted in \texttt{SurvCopBoost}. The use of a constant step-length throughout the fitting process in gradient boosting can lead to a slow convergence of the algorithm as pointed out by \cite{Zhang2022}. Since the joint survival functions set up by \texttt{SurvCopBoost} feature a large number of distribution parameters, an adaptive step-length as proposed by \cite{Zhang2022} or \cite{DauMayZha2024} would lead to considerable improvements in this area.

\section*{Acknowledgements}
The work on this article was supported by the German research foundation (DFG) through the grants KL3037/2-1, MA7304/1-1 (428239776). 

\section*{Declaration of conflicting interests}
The authors declared no potential conflicts of interest with respect to the research, authorship, and/or publication of this article.

\renewcommand\bibname{References}
\bibliographystyle{apalike}
%\bibliography{BoostBAFT_Bibliography.bib}

\pagebreak

\noindent
\begin{center}
{\bf \Large{Supplementary Material}} \\\Large{for} \\
\Large{\centering ``Boosting Distributional Copula Regression for Bivariate Right-Censored Time-to-Event Data''}
\end{center}
\begin{center}
	{\bf Contents}
\end{center}
\begin{itemize}
\item[] {\bf Part~A}: Details on implemented marginal distributions and copula functions.
\item[] {\bf Part~B}: Details on the boosting algorithm.
\item[] {\bf Part~C}: Additional results for the simulation study.

\end{itemize}

\clearpage 

\section*{Part A}
\setcounter{section}{0}\renewcommand\thesection{A\arabic{section}}
\setcounter{subsection}{0}\renewcommand\thesubsection{A\arabic{subsection}}
\setcounter{table}{0}\renewcommand{\thetable}{A\arabic{table}}
\setcounter{figure}{0}\renewcommand\thefigure{A\arabic{figure}}

\begin{table}[h!]
    \centering
    \begin{tabular}{lcccc}
    \toprule
      Distribution && $ \boldsymbol{\vartheta}^{} $ && Survival function  \\
         \midrule
    Weibull && $ \vartheta_{1}, \vartheta_{2} $ && $\exp\left( -\left( \frac{t}{\vartheta_1 } \right)^{\vartheta_2 } \right)$ \\
    \\
    Log-normal && $ \vartheta_{1}, \vartheta_{2} $ && $1-\Phi\left( \frac{\log(t) - \vartheta_1}{\vartheta_2} \right)$ \\
    \\
    Log-logistic && $ \vartheta_{1}, \vartheta_{2} $ && $1 - \frac{1}{\left( 1 + \left( \frac{t}{\vartheta_1}\right)^{-\vartheta_2 } \right) }$ \\
         \bottomrule
    \end{tabular}
    \caption{Implemented parametric distributions for right-censored time-to-event responses in \texttt{gamboostLSS}. All distribution parameters use the exponential response function, i.e.\ $\vartheta = \exp(\eta) \geq 0$ except for $\vartheta_{1}$ in the Log-normal distribution, which uses the identity link function, i.e.\ $\vartheta = \eta \in \mathbb{R}.$}\label{ImplementedAFTDistributions}
\end{table}

\begin{sidewaystable}[p]%[htbp]
 \caption{\small Details of implemented copulas for right-censored time-to-event responses. The functions $\Phi_1^{-1}(\cdot)$ and $\Phi_2(\cdot)$ denote the quantile function and CDF of the univariate and bivariate standard normal distributions, respectively.  Rotated copulas by 90, 180 and 270 degrees are respectively defined as: $C_{90} = S_2 - C(1-S_1, S_2; \vartheta^{(c)})$, $C_{180} = S_1 + S_2 - 1 + C(1-S_1, 1-S_2; \vartheta^{(c)})$ and $C_{270} = S_1 - C(S_1, 1-S_2; \vartheta^{(c)})$. 
    The term $D_1(\vartheta^{(c)}) = \int_0^{\vartheta^{(c)}} \frac{t}{\exp(t) -1 }dt$ is the Debye function and $\Phi_2$ denotes the CDF of the bivariate Gaussian distribution with correlation coefficient $\vartheta^{(c)}$. 
    }
    \centering\renewcommand{\arraystretch}{2.5}\small 
  \begin{adjustbox}{scale=0.85,center}
    \begin{tabular}{ccccc}
    \toprule
Copula & $C(S_1, S_2; \vartheta^{(c)})$ & Range of $\vartheta^{(c)}$   & Link &   Kendall's $\tau$  \\
    \midrule
     Gauss   & $\Phi_2 (\Phi_1^{-1}(S_1), \Phi_1^{-1}(S_2); \vartheta^{(c)} )$ & $\vartheta^{(c)} \in [-1, 1]$  & $\tanh^{-1}(\vartheta^{(c)})$ & $\frac{2}{\pi} \arcsin(\vartheta^{(c)})$ \\
    Clayton &  $(S_1^{-\vartheta^{(c)}} + S_2^{-\vartheta^{(c)}} - 1)^{-1/\vartheta^{(c)}}$ & $\vartheta^{(c)} \in (0, \infty)$ & $\log(\vartheta^{(c)})$ & $\frac{\vartheta^{(c)}}{\vartheta^{(c)} + 2}$ \\
    Gumbel  & $\exp\left[ - \left\{ (-\log(S_1))^{\vartheta^{(c)}} + (-\log(S_2))^{\vartheta^{(c)}} \right\}^{\tfrac{1}{\vartheta^{(c)}}} \right]$  & $\vartheta^{(c)} \in [1, \infty)$ & $\log(\vartheta^{(c)} - 1)$ & $1 - \frac{1}{\vartheta^{(c)} }$\\
 Joe & $1 - ((1 - S_1)^{\vartheta^{(c)} } + (1 - S_2)^{ \vartheta^{(c)}} - (1 - S_1)^{ \vartheta^{(c)} }  (1 - S_2)^{ \vartheta^{(c)} })^{ ( 1/\vartheta^{(c)}  ) } $  & $\vartheta^{(c)} \in [1, \infty)$ &  $\log(\vartheta^{(c)} - 1)$  & $1 + \frac{4}{ {\vartheta^{(c)} }^{2} } \int_{0}^{1} x \log(x) (1 - x)^{2 ( 1 - \vartheta^{(c)} ) / \vartheta^{(c)} } dx $ \\
   Frank   & $-{\vartheta^{(c)}}^{-1} \log\Big(1 + (\exp(-\vartheta^{(c)} S_1) - 1 ) \cdot$ & $\vartheta^{(c)} \in \mathbb{R} \setminus \{0\} $ & $\vartheta^{(c)}$ & $ 1- \frac{4}{\vartheta^{(c)}}[ 1- D_1(\vartheta^{(c)})] $ \\
     & $(\exp(-\vartheta^{(c)} S_2) - 1 ) / (\exp(-\vartheta^{(c)}) - 1) \Big)$ & & &  \\ 
         \bottomrule
    \end{tabular}
    \end{adjustbox}
    \label{ImplementedCopulas}
\end{sidewaystable}

\section*{Part B}
\setcounter{section}{0}\renewcommand\thesection{B\arabic{section}}
\setcounter{subsection}{0}\renewcommand\thesubsection{B\arabic{subsection}}
%\counterwithin*{equation}{section}
%\renewcommand\theequation{\thesection\arabic{equation}}
%\setcounter{equation}{0}\renewcommand\theequation{SB\arabic{equation}}
\setcounter{table}{0}\renewcommand{\thetable}{B\arabic{table}}
\setcounter{figure}{0}\renewcommand\thefigure{B\arabic{figure}} 
\setcounter{algorithm}{0}\renewcommand\thealgorithm{B\arabic{algorithm}}

{
\begin{algorithm}[H]%[p]
\caption{\small Two-stage, non-cyclic boosting for distributional copula regression of time-to-event responses with faster tuning of fitting iterations $\texttt{m}_{\texttt{stop}}$ by means of out-of-bag (\textit{oobag}) risk.}\label{boostingAlgorithm}
\begin{algorithmic}
\footnotesize
\Require \\
Define the base-learners $b^{(\bullet)}_{r}(x_r)$ for $r = 1, \dots, P^{(\bullet)}_{k}$, $\bullet = 1, 2, c$. 
\State Set the step-length $\texttt{s}_{\texttt{step}}  \ll 1$ as well as the (non-optimal) number of fitting iterations $\texttt{m}^{(\bullet)}_{\texttt{stop}}, \bullet = \{1, 2, c \}$. 
\State Set weights indicating the training and $\texttt{m}_{\texttt{stop}}$-tuning partitions of the sample $n_{\texttt{train}}$, $n_{\texttt{mstop}}$.
\State Set stabilisation to be applied to the negative gradient vector ($L_2$, median absolute deviation or none).
\For{$\bullet = \{ 1,2 \}$}
\State (1) Initialise all predictors $\hat{\eta}^{(\bullet)}_{k}$ corresponding to $\vartheta_k^{(\bullet)} \in \boldsymbol{\vartheta}^{(\bullet)}$ with offset values $\hat{\eta}^{(\bullet)}_{k, [0]}$. 
\For{$m = 1, \dots, \texttt{m}_{\texttt{stop}}^{(\bullet)}$}
\For{$k = 1, \dots, K_{\bullet}$ in $\vartheta_k^{(\bullet)} \in \boldsymbol{\vartheta}^{(\bullet)}$ }
\State (a) Evaluate the parameter-specific negative gradient vector $\boldsymbol{-g}^{(\bullet)}_{k, [m]}$ 
    \begin{equation*}
\boldsymbol{-g}^{(\bullet)}_{k, [m]} =
\left( \boldsymbol{-g}^{(\bullet)}_{k, [m]} (\boldsymbol{x}_i)  \right)_{i = 1, \dots, n_{\text{train}}} = 
- \left( \left.\frac{ \partial \omega\left( \boldsymbol{y}_i, \boldsymbol{\hat{\eta}}^{(\bullet)}_{i} \right)  }{ \partial \eta^{(\bullet)}_{k} }\right\vert_{ \boldsymbol{\hat{\eta}^{(\bullet)} } = \boldsymbol{\hat{\eta} }^{(\bullet)}_{[m-1] }(\boldsymbol{x}_i) }  \right)_{i = 1, \dots, n_{\text{train} } }.
    \end{equation*}
\State (b) Fit $\boldsymbol{-g}^{(\bullet)}_{k, [m]}$ to each parameter-specific base-learner $b^{(\bullet)}_{k,j}(x_j)$.
\State (c) Select the best-fitting base-learner $\hat{b}^{(\bullet)}_{k,j^{\star}}$ via residual sum of squares criterion.
\begin{equation*}
j^{\star} = \argmin_{j \in 1, \dots P_k^{(\bullet)} } \sum_{i = 1}^{ n_{ \text{train} } } \left( -g^{(\bullet)}_{k, [m]}(\boldsymbol{x}_i) -  \hat{b}^{(\bullet)}_{k,j^{}}(x_{i})  \right)^2.
\end{equation*}
\State (d) Compute loss reduction of a weak update using $\hat{b}^{(\bullet)}_{k,j^{\star}}$.
\begin{equation*}
 \Delta \omega_{\vartheta^{(\bullet)}_k} = \sum_{i = 1}^{ n_{ \text{train} } } \omega\left( \boldsymbol{y}_i; \hat{\eta}^{(\bullet)}_k + \texttt{s}_{\texttt{step}}   \hat{b}^{(\bullet)}_{k,j^{\star}}( x_{ij^{\star}} ) \right).
\end{equation*}
\EndFor
\State (2) Update the parameter with highest loss reduction $\vartheta^{(\bullet)^\star }_{k} = \argmin_{ \vartheta^{(\bullet) }_{k} \in \boldsymbol{\vartheta} } \left( \Delta \omega_{\vartheta^{(\bullet)}_k} \right)$:
\begin{equation*}
    \hat{\eta}^{(\bullet)*}_{k, [m]}(\boldsymbol{x}_i) =
     \hat{\eta}^{(\bullet)*}_{k, [m-1]}(\boldsymbol{x}_i) + \texttt{s}_{\texttt{step}}  \cdot \hat{b}^{(\bullet)}_{k,j^{\star}}( x_{ij^{\star}} ).
\end{equation*}

\State (3) For the remaining parameters $\vartheta^{(\bullet)}_k \neq \vartheta^{(\bullet)^\star }_k$, set $\hat{\eta}^{(\bullet)}_{k, [m]} (\boldsymbol{x}_i) = \hat{\eta}^{(\bullet)}_{k, [m-1]} (\boldsymbol{x}_i) $.
\State (4) Compute the out-of-bag risk at iteration $[m]$ : 
\begin{equation*}
\text{risk}^{(\bullet)}_{\text{oobag}, [m]} = \sum_{i = 1}^{n_{ \texttt{mstop} }}  \hat{\omega}\left( \left. \boldsymbol{y}_i; \boldsymbol{\hat{\eta}}^{(\bullet)}_{i}  \right\vert_{\hat{\eta}^{(\bullet)} = \hat{\eta}^{(\bullet)}_{[m]} (\boldsymbol{x}_i) }   \right)  .
\end{equation*}
\EndFor
\State (5) Determine $\texttt{m}^{\texttt{opt} (\bullet) }_{\texttt{stop}}$ by means of the out-of-bag-risk: 
\begin{equation*}
     \texttt{m}^{ \texttt{opt} (\bullet) }_{\texttt{stop}} = \argmin_{m \in 1, \dots , \texttt{m}^{(\bullet)}_{\texttt{stop}} } \text{risk}_{\text{oobag}, [m]}.
\end{equation*}
\EndFor
\State (6) Compute $\hat{S}_\bullet\left(y_{\bullet i}; \boldsymbol{\hat\vartheta}_{i}^{(\bullet)} \right)$, $\hat{f}_\bullet\left(y_{\bullet i}; \boldsymbol{\hat\vartheta}_{i}^{(\bullet)} \right)$ using $\texttt{m}^{\texttt{opt} (\bullet) }_{\texttt{stop}}$, $\bullet = \{1,2\}$. Plug them into the loss of Equation~\eqref{BIVAFTLOGLIK}. 
\State (7) Conduct steps (1)-(5) using the loss of Equation~\eqref{BIVAFTLOGLIK} with $\bullet = c$ in order to determine $\texttt{m}^{\texttt{opt} (c) }_{\texttt{stop}}$.
\end{algorithmic}
\end{algorithm}}

\noindent Note that during the first for-loop the loss function in steps (1)-(5) in Algorithm~\ref{boostingAlgorithm} is set to the negative log-likelihood of univariate right-censored responses 
\begin{equation*}
   \omega_i =  -\ell_i = - \left( \delta_{\bullet i} \log\left( f_\bullet\left( y_{\bullet i}; \boldsymbol{\vartheta}_{i}^{(\bullet)} \right) \right) + 
   (1 - \delta_{\bullet i}) \log\left( 
   S_\bullet \left( y_{\bullet i}; \boldsymbol{\vartheta}_{i}^{(\bullet)} \right)
   \right)
   \right), \quad \bullet = \{1,2\},
\end{equation*}
whereas for the remainder of the steps it is set to the negative log-likelihood that corresponds to Equation~\eqref{BIVAFTLOGLIK}.

\subsection{Fitting bivariate distributional copula regression models for right-censored data using \texttt{SurvCopBoost} in \textsf{R}}\label{UsingSurvCopBoost}

We briefly illustrate how to use the \textsf{R} routine \texttt{SurvCopBoost} which implements Algorithm~\ref{boostingAlgorithm}. The function uses syntax similar to that of \texttt{mboost}, \texttt{gamboostLSS} and other regression routines:
\begin{verbatim}
## All covariates enter the model of margin 1
Formula_Margin1 <- list(mu = cbind(time1, cens1) ~ .,
                        sigma = cbind(time1, cens1) ~ .)

## All covariates enter the model of margin 2
Formula_Margin2 <- list(mu = cbind(time1, cens1) ~ .,
                        sigma = cbind(time1,cens1) ~ .)

## All covariates enter the model of the copula parameter
Dependence_Formula <- cbind(SURV1, PDF1, delta1, 
                            SURV2, PDF2, delta2) ~ .

## Construct list of formulas
formula_list <- list(Formula_Margin1, 
                     Formula_Margin2,
                     Dependence_Formula)

## Fit the model, consider 1000 iterations for each sub-model
Fit <- SurvCopBoost(formulas = formula_list, 
                    marings = c("WEIBULL", "LOGLOGISTIC"), 
                    copula = c("GUMBEL"), 
                    response_1 = resp1, response_2 = resp2, data = dat, 
                    mstops = c(1000, 1000, 1000),
                    oobag_weights = boost_weights, 
                    s_step = 0.1, stabilization = "L2")
\end{verbatim}
The argument \texttt{formulas} requires a list with three entries that indicate the formulas used for fitting the model of the two margins as well as the dependence parameter $\vartheta^{(c)}$. The marginal distributions are specified in the argument \texttt{margins}, which supports the entries \texttt{WEIBULL}, \texttt{LOGNORMAL}, and \texttt{LOGLOGISTIC}. The copula function is determined by the argument \texttt{copula}. Rotated copulas are specified by entering the degrees of rotation, e.g.\ \texttt{GUMBEL270} for a Gumbel copula by 270$\degree$. The arguments \texttt{response\_1} and \texttt{response\_2} are data frames of dimension $n \times 2$, where the first column is the time variable and the second column is the censoring indicator parsed as a binary variable. The explanatory variables are provided in the \texttt{data} argument. Note that \texttt{data} should not contain the time variables and censoring indicators. A vector of length $n$ consisting only of binary entries must be supplied for \texttt{oobag\_weights}. This determines the observations used for fitting and for the tuning of \texttt{m\_{stop}}. The out-of-bag risk is computed on the observations with weight equal to zero. Lastly, the arguments \texttt{mstops}, \texttt{s\_step} and \texttt{stabilization} specify the hyperparameters of the boosting algorithm. \\

\noindent The formula of the dependence parameter declared in \texttt{Dependence\_Formula} requires the structure with the provided names (\texttt{SURV1}, \texttt{PDF1}, \texttt{delta1}, etc.). These objects denote the survival function, probability density function and the censoring indicator of each margin, respectively. The marginal survival functions and probability density functions are computed internally after boosting each margin as described in Step (6) of Algorithm~\ref{boostingAlgorithm}. The output of \texttt{SurvCopBoost} is a list which contains the individual sub-models of the margins and the dependence parameter. These objects can then be used with typical convenience functions such as \texttt{predict}, \texttt{plot}, \texttt{coef}, and \texttt{summary} from the \texttt{gamboostLSS} package.

\pagebreak
\section*{Part C}
\setcounter{section}{0}\renewcommand\thesection{C\arabic{section}}
\setcounter{subsection}{0}\renewcommand\thesubsection{C\arabic{subsection}}
%\counterwithin*{equation}{section}
%\renewcommand\theequation{\thesection\arabic{equation}}
%\setcounter{equation}{0}\renewcommand\theequation{SB\arabic{equation}}
\setcounter{table}{0}\renewcommand{\thetable}{C\arabic{table}}
\setcounter{figure}{0}\renewcommand\thefigure{C\arabic{figure}}

\begin{figure}[h!]
    \centering
    \includegraphics[scale = 0.65]{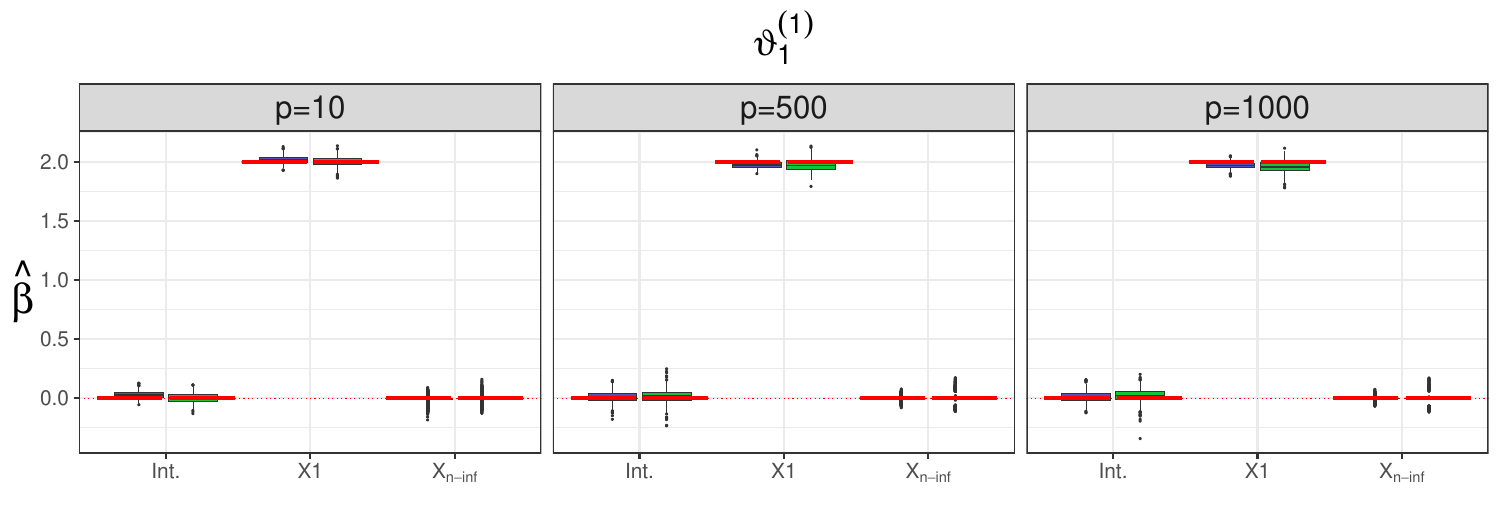}\\
    \includegraphics[scale = 0.65]{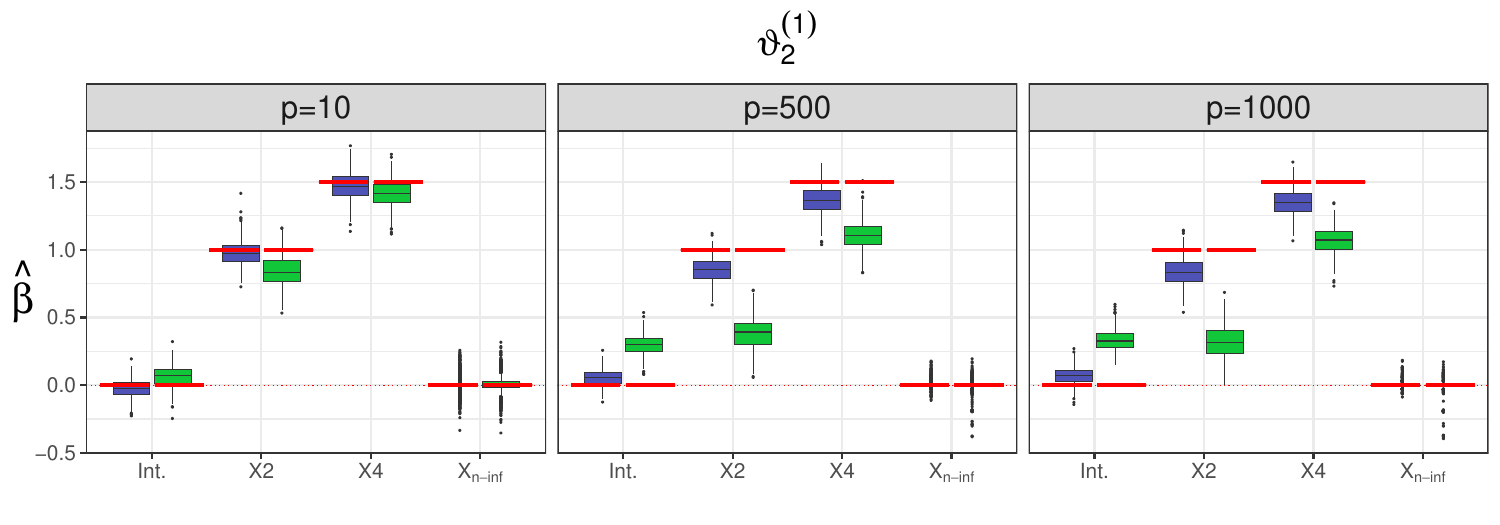}\\
    \includegraphics[scale = 0.65]{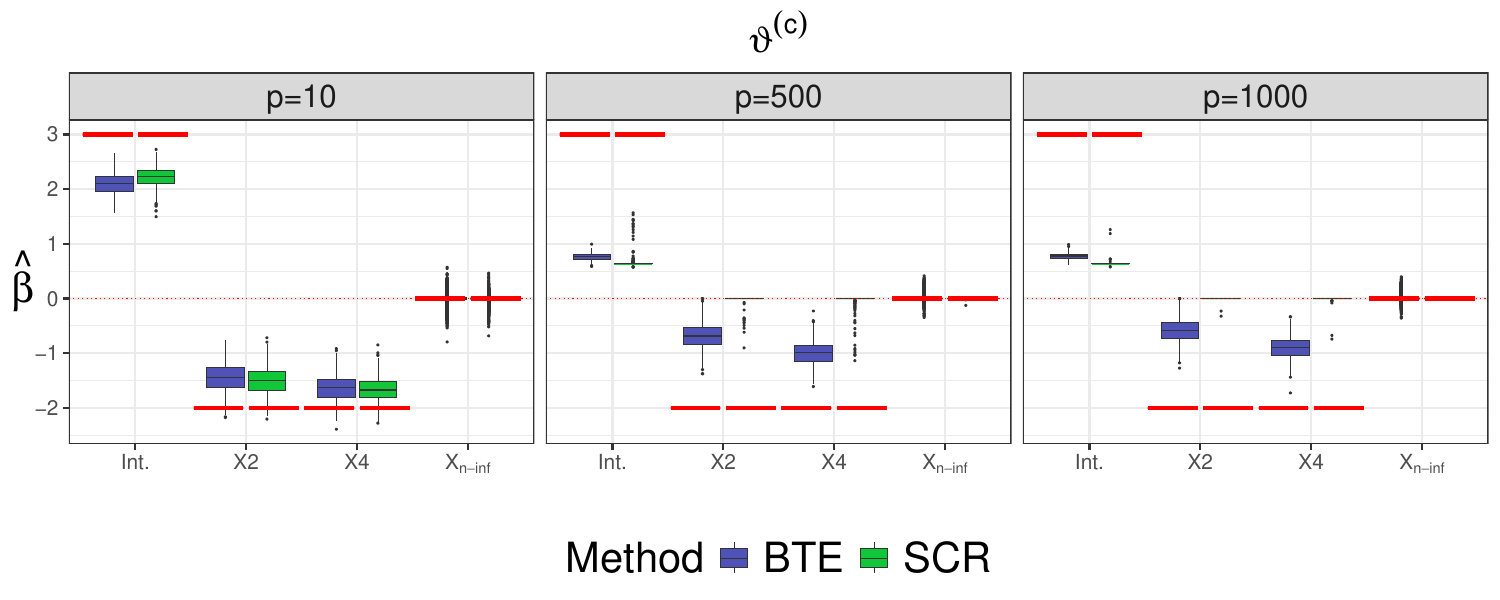}
    \caption{Simulation study 1 (SCR responses). Estimated coefficients of the copula model across distribution parameters, number of potential covariates using BTE and SCR estimation methods based on 500 independent replications. Thick red lines denote true values.}
    \label{SIM2_LINDGP_TOEP}
\end{figure}

\begin{figure}[h!]
    \centering
 \includegraphics[scale = 0.65]{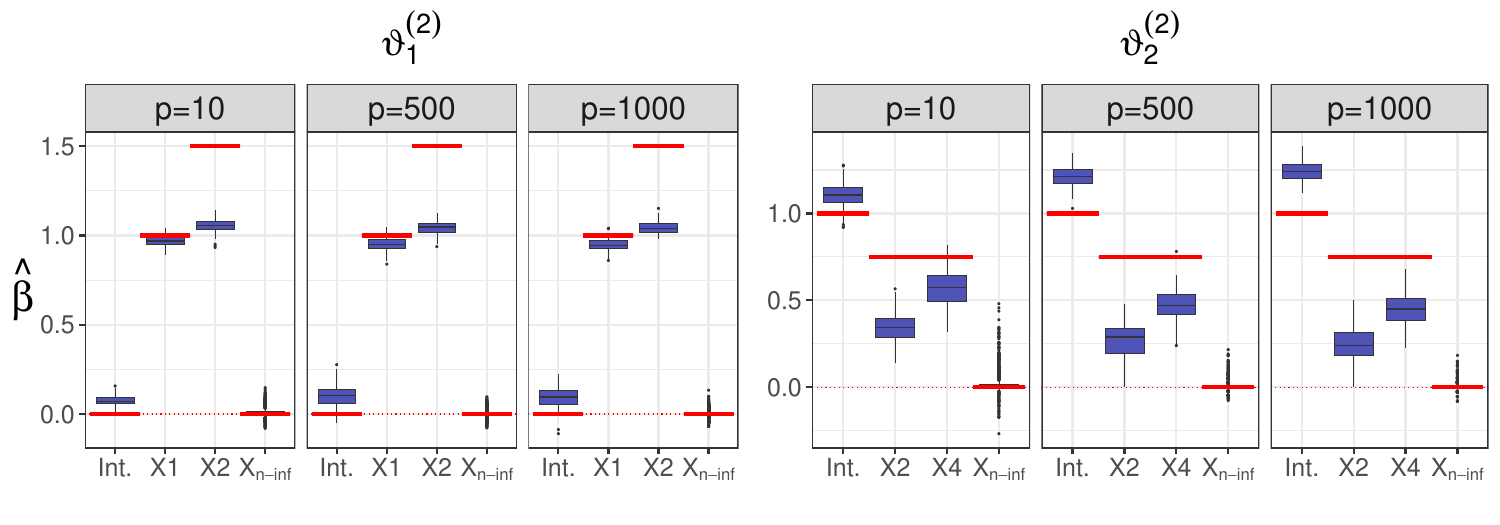}\\
    \caption{Simulation study 1 (SCR responses). Estimated coefficients of the copula model in the margin corresponding to the terminal event ($T_{2}$) across distribution parameters and number of potential covariates using 500 independent replications. }
    \label{SIM2_LINDGP_M2ONLY_TOEP}
\end{figure}

\begin{figure}[h!]
    \centering
    \includegraphics[scale = 0.5]{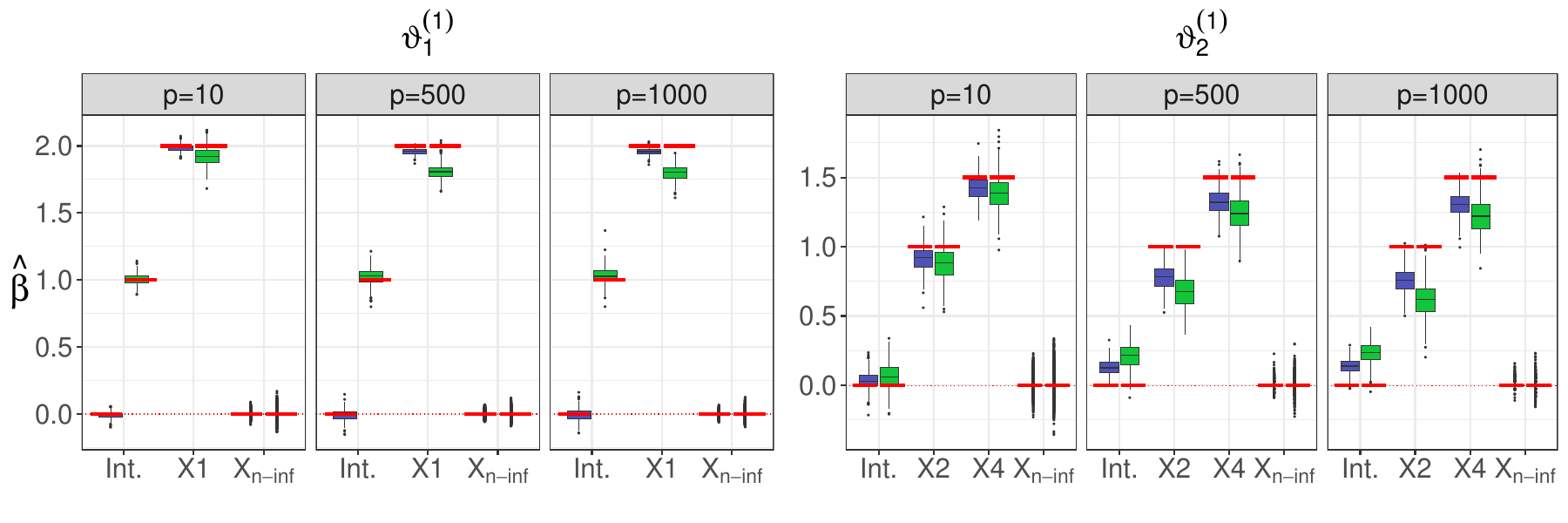}\\
    \includegraphics[scale = 0.5]{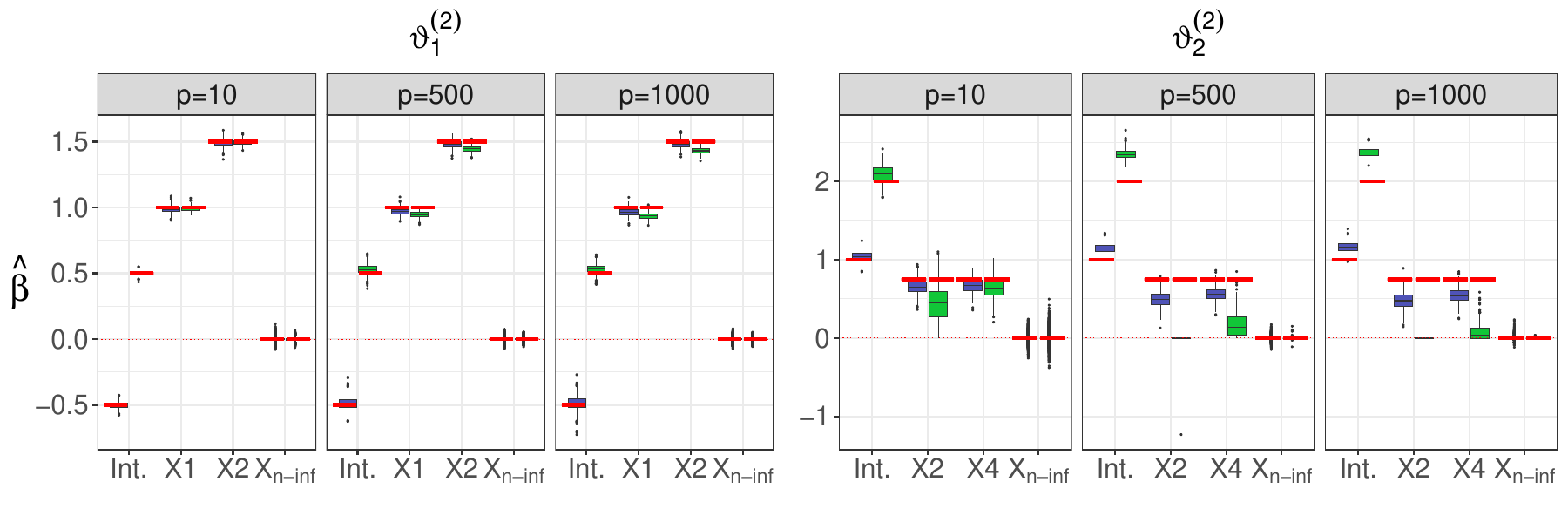}\\
    \includegraphics[scale = 0.5]{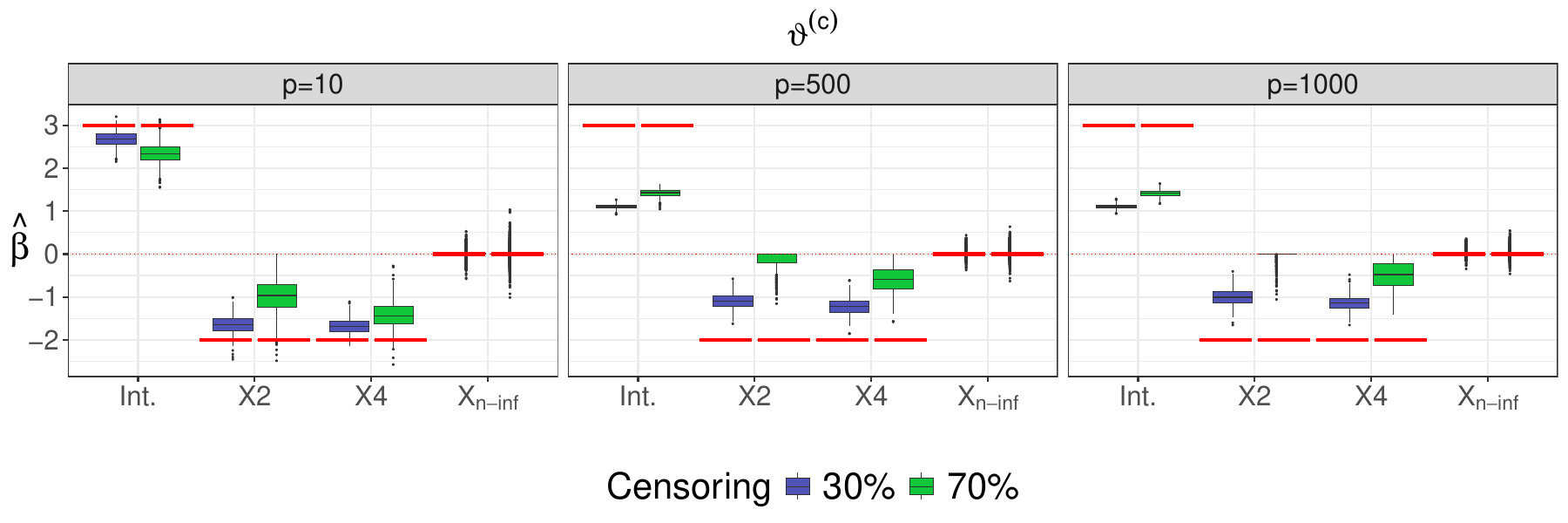}
    \caption{Simulation study 2 linear DGP. Estimated coefficients of the copula model across distribution parameters, number of potential covariates and censoring rates using 500 independent replications. Thick red lines denote true values.}
    \label{SIM1_LINDGP_TOEP}
\end{figure}

\begin{figure}[h!]
    \centering
    \includegraphics[scale = 0.45]{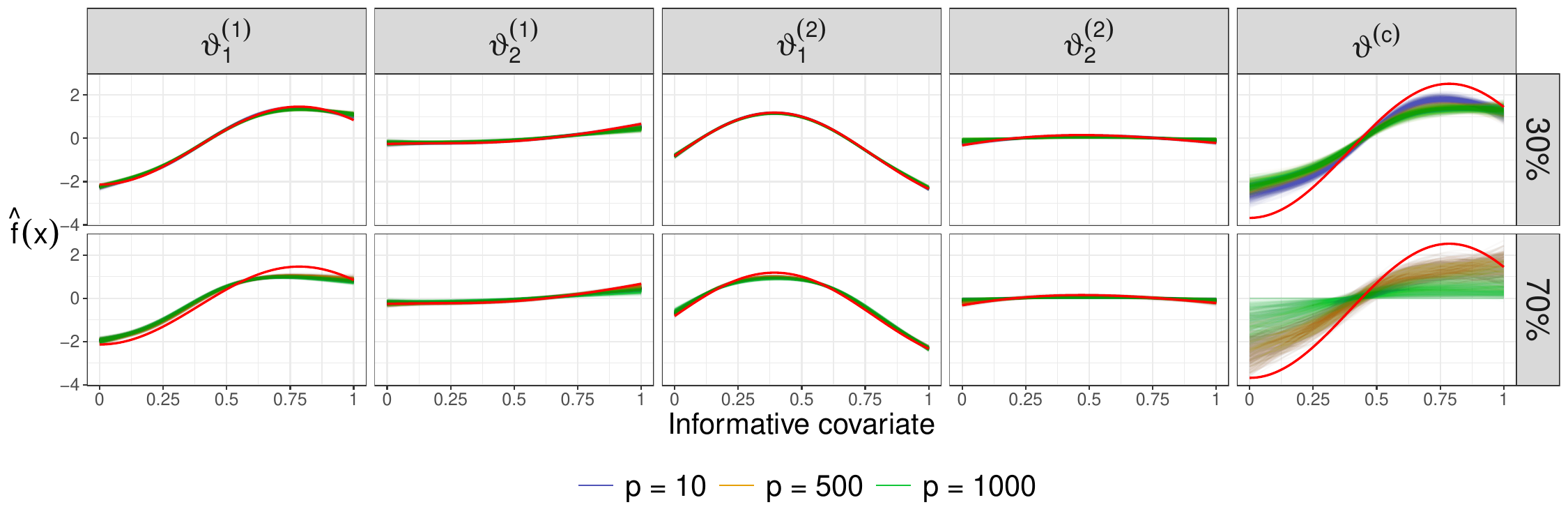}
      \caption{Simulation study 2 non-linear DGP. Estimated non-linear effects of the copula model across distribution parameters, number of potential covariates and censoring rates using 500 independent replications. Thick red lines denote the true non-linear functions.}
    \label{SIM1_NLDGP_TOEP}
\end{figure}

{\footnotesize
\begin{table}[!h]
    \centering
    \caption{\small Simulation study 1 (SCR responses). Performance metrics for the simulation studies for the copula models using BTE and SCR estimation, as well as independent univariate Cox models ($Cox$). Values are mean scores from the 500 independent replicates (each evaluated on the test dataset), whereas
    parentheses show the respective standard deviations. }
    \resizebox{\columnwidth}{!}{%
    \begin{tabular}{l l c c c c c c}
    \toprule
    %  \\
    % & Model &  \multicolumn{6}{c}{}   \\
      & Model & &  $p_1 = 10$ & &  $p_2 = 500$ && $p_3 = 1000$  \\
             %& \\
              %& \\  
         \midrule
             \rowcolor{Gainsboro!60}
     log-score      & $BTE$ & & $\phantom{0}842.821\ (38.263)$ & & $\phantom{0}884.854 \ (40.024)$ && $\phantom{0}894.886\ (39.025)$ \\
               & $SCR$ & & $\phantom{0}829.837\ (37.451)$ & & $\phantom{0}932.535\ (35.847)$ && $\phantom{0}939.632 \ (37.148)$  \\
                  \rowcolor{Gainsboro!60}
               & $Ind$ & & $1257.613\ (43.103)$ & & $1299.857\ (45.343)$ && $1312.84\ (43.517)$\\
               %\\
      %\midrule
      \midrule
      \rowcolor{Gainsboro!60}
     IBS ($T_{1}$)  & $BTE$ & & $0.180\ (0.205)$ & & $0.182\ (0.213)$ &&  $0.175\ (0.195)$ \\
                    & $SCR$ & & $0.179\ (0.203)$ & & $0.188\ (0.219)$ &&  $0.181\ (0.199)$\\
                    \rowcolor{Gainsboro!60}
                    & $Cox$ & & $0.458\ (0.142)$ && $0.465\ (0.147)$ && $0.458\ (0.137)$ \\
                   % \\
         \midrule
       \rowcolor{Gainsboro!60}
     IBS ($T_{2}$)  & $BTE$ & & $0.198\ (0.231)$ && $0.190 \ (0.225)$ && $0.182\ (0.214)$ \\
                    & $SCR$ & & $0.198\ (0.231)$ && $0.190 \ (0.225)$ && $0.182\ (0.214)$ \\
                    \rowcolor{Gainsboro!60}
                    & $Cox$ & & $0.376\ (0.118)$ && $0.371\ (0.116)$ && $0.354\ (0.109)$ \\
                  %  \\
         \midrule
             \rowcolor{Gainsboro!60}
     ISE ($T_{1}$)  & $BTE$ & & $0.002 \ (0.001)$ && $0.004 \ (0.001)$ && $0.005\ (0.002)$\\
                    & $SCR$ & & $0.003 \ (0.001)$ && $0.015 \ (0.004)$ && $0.017\ (0.005)$\\
                    \rowcolor{Gainsboro!60}
                    & $Cox$ & & $1.966\ (0.127)$ && $1.979\ (0.167)$ && $1.979\ (0.155)$\\
                %    \\
         \midrule
             \rowcolor{Gainsboro!60}
     ISE ($T_{2}$)  & $BTE$ & & $0.003\ (0.001)$ && $0.009\ (0.002)$ && $0.011\ (0.003)$ \\
                    & $SCR$ & & $0.003\ (0.001)$ && $0.009\ (0.002)$ && $0.011\ (0.003)$ \\
                    \rowcolor{Gainsboro!60}
                    & $Cox$ & & $1.736\ (0.094)$  && $1.728\ (0.129)$  && $1.731\ (0.123)$ \\
              %      \\
         \midrule
             \rowcolor{Gainsboro!60}
     IAE ($T_{1}$)  & $BTE$ && $0.063\ (0.018)$ && $0.085\ (0.014)$ && $0.092\ (0.015)$\\
                    & $SCR$ && $0.070\ (0.018)$ && $0.180\ (0.023)$ && $0.199\ (0.027)$\\
                    \rowcolor{Gainsboro!60}
                    & $Cox$ && $2.921\ (0.126)$ && $2.938\ (0.157)$ && $2.940\ (0.148)$\\
             %       \\
         \midrule
             \rowcolor{Gainsboro!60}
     IAE ($T_{2}$)  & $BTE$ && $0.073\ (0.018)$ && $0.143\ (0.020)$ && $0.160\ (0.023)$\\
                    & $SCR$ && $0.073\ (0.018)$ && $0.143\ (0.020)$ && $0.160\ (0.023)$\\
                    \rowcolor{Gainsboro!60}
                    & $Cox$ && $2.468\ (0.089)$ && $2.459\ (0.113)$ && $2.464\ (0.107)$\\
              %      \\
         \midrule
             \rowcolor{Gainsboro!60}
     C-Index ($T_{1}$)  & $BTE$ & & $0.824\ (0.008)$ && $0.825\ (0.008)$ && $0.824\ (0.008)$\\
                        & $SCR$ & & $0.824\ (0.008)$ && $0.824\ (0.008)$ && $0.824\ (0.008)$\\
                    \rowcolor{Gainsboro!60}
                        & $Cox$ & & $0.823\ (0.008)$ && $0.825\ (0.008)$&& $0.824\ (0.008)$ \\
              %      \\
         \midrule
             \rowcolor{Gainsboro!60}
     C-Index ($T_{2}$)  & $BTE$ & & $0.862\ (0.008)$ && $0.861 \ (0.008)$ && $0.861\ (0.008)$ \\
                        & $SCR$ & & $0.862\ (0.008)$ && $0.861 \ (0.008)$ && $0.861\ (0.008)$ \\
                    \rowcolor{Gainsboro!60}
                    & $Cox$ & & $0.861\ (0.008)$ && $0.861\ (0.008)$ && $0.860\ (0.008)$ \\
             %       \\
       %  \midrule
      \midrule
      \midrule
                                %\rowcolor{Gainsboro!60}
    \multicolumn{8}{l}{Gumbel copula with Kendall's $\tau$ with range within $[0.187;\ 0.922]$.  } \\
    \\
    \multicolumn{8}{l}{Gradients stabilised using $L_2$ norm, step-length $\texttt{s}_{\texttt{step}} = 0.1$. $n_{\text{train}} = 1000$, $n_{\text{test}} = 1000$, $n_{\texttt{mstop}} = 1000$. } \\
\bottomrule
    \end{tabular}
    }\label{SimulationsSCRMetrics_TOEPLITZ}
\end{table}}

\begin{table}[h!]
    \centering
    \caption{\small Simulation study 1 (SCR responses). True positive rates (TPR) and false positive rates (FPR) for the copula models using BTE and SCR estimation for each distribution parameter as well as independent univariate Cox models ($Cox$) for each margin.  
    Values are averages over $500$ independent datasets.
    }
    %\resizebox{\columnwidth}{!}{%
    \begin{tabular}{l ccccccccc } % 20 columns
    \toprule
%           &   & \multicolumn{8}{c}{$(1)$} & & & \multicolumn{8}{c}{ $(2)$ }  \\
    
%  & & \multicolumn{8}{c}{ $\approx 44\%$ censoring } \\
%  \\
 & & \multicolumn{2}{c}{ $p_1 = 10$ } & & \multicolumn{2}{c}{ $p_2 = 500$ } & & \multicolumn{2}{c}{ $p_3 = 1000$ } \\
              \\
   & &  TPR & FPR & & TPR & FPR & & TPR & FPR \\
              %& \\  
         \midrule
         %  \multirow{15}{*}{\STAB{\rotatebox[origin=c]{90}{rota}}} \\
  %   & &    \multicolumn{18}{c}{Linear DGP}  \\
        % \midrule
   \multicolumn{10}{l}{Copula model $(BTE)$ }  \\
   \\
  \rowcolor{Gainsboro!60}
$\vartheta_{1}^{(1)}$ &  & $1$ & $0.296$ & & $1$ & $0.029$ & & $1$ & $0.015$ \\
$\vartheta_{2}^{(1)}$ &  & $1$ & $0.173$ & & $1$ & $0.001$ & & $1$ & $0.000$ \\
\\
 \rowcolor{Gainsboro!60}
$\vartheta_{1}^{(2)}$ &  & $1$ & $0.253$ & & $1$ & $0.040$ & & $1$ & $0.021$ \\
$\vartheta_{2}^{(2)}$ &  & $1$ & $0.353$ & & $0.990$ & $0.003$ & & $0.970$ & $0.001$ \\
\\
 \rowcolor{Gainsboro!60}
$\vartheta_{}^{(c)}$ &  & $1$ & $0.291$ & & $0.998$ & $0.054$ &  & $0.997$ & $0.027$ \\
\midrule
   \multicolumn{10}{l}{Copula model $(SCR)$ }  \\
   \\
  \rowcolor{Gainsboro!60}
$\vartheta_{1}^{(1)}$ &  & $1$ & $0.145$ & & $1$ & $0.004$ & & $1$ & $0.002$ \\
$\vartheta_{2}^{(1)}$ &  & $1$ & $0.353$ & & $1$ & $0.002$ & & $0.996$ & $0.000$ \\
\\
 \rowcolor{Gainsboro!60}
$\vartheta_{1}^{(2)}$ &  & $1$ & $0.253$ & & $1$ & $0.040$ & & $1$ & $0.021$ \\
$\vartheta_{2}^{(2)}$ &  & $1$ & $0.353$ & & $0.990$ & $0.003$ & & $0.970$ & $0.001$ \\
\\
 \rowcolor{Gainsboro!60}
$\vartheta_{}^{(c)}$ &  & $1$ & $0.141$ & & $0.055$ & $0.000$ & & $0.009$ & $0.000$ \\
\midrule
 \multicolumn{10}{l}{Cox models $(Cox)$ }  \\ %%  UNIVARIATE MODEL!!!
 \\
 \rowcolor{Gainsboro!60}
Margin 1 &  & $0.791$ & $0.223$ & & $0.475$ & $0.034$ & & $0.431$ & $0.021$ \\
Margin 2 &  & $0.913$ & $0.201$ & & $0.751$ & $0.039$ & & $0.721$ & $0.025$ \\
\midrule
\midrule
   \multicolumn{10}{l}{Gumbel copula with Kendall's $\tau$ range within $[0.187;\ 0.922]$.  } \\
    \multicolumn{10}{l}{Gradients stabilised using $L_2$ norm, step-length $\texttt{s}_{\texttt{step}} = 0.1$.}\\
    \multicolumn{10}{l}{$n_{\text{train}} = 1000$, $n_{\text{test}} = 1000$, $n_{\texttt{mstop}} = 1000$. } \\
\bottomrule
    \end{tabular}
 %   }
    \label{Simulation2SelectionRatesSCR_TOEP}
\end{table}

{\footnotesize
\begin{table}[t!]
    \centering
        \caption{\small Simulation study 2. Performance metrics for the simulation studies for the copula ($Cop$), independent models ($Ind$), and Cox models ($Cox$), $\star$ identifies the non-linear DGP. Values are mean scores from the 500 independent replicates (each evaluated on  the test dataset), whereas
    parentheses show the respective standard deviations. }
    \resizebox{\columnwidth}{!}{%
    \begin{tabular}{l l c c c c c c c c c }
    \toprule
          &       &  &  \multicolumn{3}{c}{$(1)$} & & & \multicolumn{3}{c}{$(2)$}   \\
      \\
     & Model & & \multicolumn{3}{c}{30\% censoring} & & & \multicolumn{3}{c}{70\% censoring}  \\
      &  & &  $p_1 = 10$ & $p_2 = 500$ & $p_3 = 1000$ & & & $p_1 = 10$ & $p_2 = 500$ & $p_3 = 1000$ \\
             %& \\
              %& \\  
         \midrule
             \rowcolor{Gainsboro!60}
     log-score     & $Cop$ & & $1065.961\ (39.221)$ & $1105.877 \ (40.192)$ & $1112.354 \ (41.310)$ & & & $774.550 \ (35.595)$ & $825.408 \ (35.945)$ & $837.167 \ (36.137)$ \\
             & $Ind$ & & $1462.413 \ (40.854)$ & $1476.543 \ (41.899)$ & $1479.465 \ (41.955)$ & & & $922.435 \ (38.587)$ & $951.135 \ (38.072)$ & $959.964 \ (37.356)$ \\
                               \\
            \rowcolor{Gainsboro!60}
 & $Cop\ \star$ & & $1103.680\ (46.274)$ & $1155.296\ (49.054)$ & $1165.732\ (44.215)$ &&& $166.023\ (30.444)$ & $189.372\ (30.626)$ & $193.941\ (30.857)$ \\
& $Ind\ \star$ & & $1383.551\ (49.858)$ & $1395.217\ (52.250)$ & $1400.523\ (46.327)$ &&& $183.126\ (31.038)$ & $199.275\ (31.044)$ & $203.357\ (31.048)$ \\
      \midrule
      \midrule
      \rowcolor{Gainsboro!60}
 IBS ($T_{1}$)  & $Cop$ &   & $0.158 \ (0.198)$ & $0.162 \ (0.198)$ & $0.139 \ (0.172)$ &&& $0.212 \ (0.181)$ & $0.209 \ (0.181)$ & $0.197 \ (0.168)$ \\
                & $Cox$ & & $0.402 \ (0.157)$ & $0.408 \ (0.155)$ & $0.388 \ (0.130)$  &&&  $0.488 \ (0.151)$ & $0.468 \ (0.147)$ & $0.462 \ (0.130)$ \\
                                               \\
          \rowcolor{Gainsboro!60}
    & $Cop\ \star$ & & $0.249\ (0.254)$ & $0.238\ (0.242)$ & $0.243\ (0.251)$ & & & $0.276\ (0.243)$ & $0.263\ (0.235)$ & $0.260\ (0.236)$ \\
    & $Cox\ \star$ & & $0.315\ (0.287)$ & $0.305\ (0.284)$ & $0.315\ (0.294)$ & & & $0.258\ (0.253)$ & $0.249\ (0.246)$ & $0.245\ (0.294)$ \\
         \midrule
      \rowcolor{Gainsboro!60}
IBS ($T_{2}$)  & $Cop$ & & $0.154 \ (0.220)$ & $0.190 \ (0.257)$ & $0.183 \ (0.248)$ &&& $0.109 \ (0.188)$ & $0.097 \ (0.176)$ & $0.107 \ (0.185)$ \\
                & $Cox$   & & $0.464 \ (0.159)$ & $0.490 \ (0.182)$ & $0.487 \ (0.177)$ &&&   $0.443 \ (0.130)$ & $0.435 \ (0.131)$ & $0.435 \ (0.130)$ \\
                                               \\
    \rowcolor{Gainsboro!60}
        & $Cop\ \star$ && $0.249\ (0.305)$ & $0.252\ (0.306)$ & $0.235\ (0.292)$ &&& $0.187\ (0.230)$ & $0.191\ (0.226)$ & $0.190\ (0.229)$ \\
        & $Cox\ \star$ && $0.397\ (0.106)$ & $0.393\ (0.106)$ & $0.387\ (0.101)$ &&& $0.245\ (0.056)$ & $0.239\ (0.059)$ & $0.240\ (0.058)$ \\
        \midrule
        \rowcolor{Gainsboro!60}
ISE ($T_{1}$)  & $Cop$   & & $0.002\ (0.001)$ & $0.006\ (0.002)$ & $0.007\ (0.002)$     &&& $0.005\ (0.003)$ & $0.017\ (0.005)$ & $0.019\ (0.005)$ \\
                    & $Cox$   & & $2.617\ (0.268)$ & $2.639\ (0.280)$ & $2.633\ (0.302)$ &&& $1.180\ (0.078)$ & $1.125\ (0.101)$ & $1.128\ (0.302)$\\
                                               \\
          \rowcolor{Gainsboro!60}
                & $Cop\ \star$   & & $0.007\ (0.003)$ & $0.014\ (0.004)$ & $0.015\ (0.004)$ &&& $0.001\ (0.000)$ & $0.003\ (0.001)$ & $0.003\ (0.001)$ \\
                & $Cox\ \star$ & & $0.703\ (0.047)$   & $0.692\ (0.048)$ & $0.694\ (0.047)$ &&& $0.042\ (0.004)$ & $0.041\ (0.004)$ & $0.041\ (0.047)$ \\
         \midrule
      \rowcolor{Gainsboro!60}
ISE ($T_{2}$) & $Cop$    & & $0.002\ (0.001)$ & $0.005\ (0.002)$ & $0.006\ (0.002)$ & & & $0.002\ (0.001)$ & $0.016\ (0.004)$ & $0.019\ (0.004)$ \\
                   & $Cox$    & & $3.157\ (0.297)$ & $3.217\ (0.343)$ & $3.199\ (0.349)$ & & & $2.263\ (0.073)$ & $2.259\ (0.078)$ & $2.257\ (0.078)$ \\
                                               \\
    \rowcolor{Gainsboro!60}
              & $Cop\ \star$   & & $0.003\ (0.001)$ & $0.007\ (0.002)$ & $0.007\ (0.002)$ &&& $0.001\ (0.000)$ & $0.003\ (0.001)$ & $0.003\ (0.001)$ \\
                & $Cox\ \star$ & & $0.710\ (0.057)$ & $0.701\ (0.058)$ & $0.705\ (0.055)$ &&& $0.077\ (0.008)$ & $0.071\ (0.006)$ & $0.071\ (0.006)$ \\
                \midrule
        \rowcolor{Gainsboro!60}
IAE ($T_{1}$)  & $Cop$   & & $0.057\ (0.017)$ & $0.110\ (0.019)$ & $0.120\ (0.019)$ & & & $0.109\ (0.032)$ & $0.216\ (0.036)$ & $0.233\ (0.036)$ \\
                    & $Cox$ & & $3.865 \ (0.307)$  & $3.903\ (0.306)$ & $3.893\ (0.330)$ & & & $2.045\ (0.074)$ & $1.994\ (0.096)$ & $1.998\ (0.330)$ \\
                                               \\
          \rowcolor{Gainsboro!60}
                & $Cop\ \star$   & & $0.130\ (0.025)$ & $0.186\ (0.028)$ & $0.193\ (0.028)$ &&& $0.024\ (0.005)$ & $0.036\ (0.006)$ & $0.038\ (0.007)$ \\
                & $Cox\ \star$ & & $1.685\ (0.068)$   & $1.680\ (0.070)$ & $1.685\ (0.070)$ &&& $0.164\ (0.009)$ & $0.165\ (0.009)$ & $0.164\ (0.070)$ \\
         \midrule
      \rowcolor{Gainsboro!60}
IAE ($T_{2}$)  & $Cop$  & & $0.058\ (0.014)$ & $0.106\ (0.016)$ & $0.115\ (0.016)$ & & & $0.052\ (0.013)$ & $0.145\ (0.021)$ & $0.162\ (0.019)$ \\
                    & $Cox$  & & $4.299\ (0.301)$ & $4.359\ (0.328)$ & $4.335\ (0.338)$ & & & $2.590\ (0.072)$ & $2.586\ (0.075)$ & $2.588\ (0.075)$ \\
                                               \\
    \rowcolor{Gainsboro!60}
              & $Cop\ \star$    & & $0.088\ (0.016)$ & $0.133\ (0.018)$ & $0.140\ (0.018)$ &&& $0.021\ (0.003)$ & $0.037\ (0.005)$ & $0.040\ (0.006)$ \\
                & $Cox\ \star$ &  & $1.596\ (0.074)$ & $1.589\ (0.074)$ & $1.595\ (0.069)$ &&& $0.216\ (0.010)$ & $0.213\ (0.011)$ & $0.212\ (0.010)$ \\
        \midrule
         \rowcolor{Gainsboro!60}
C-Index ($T_{1}$)  & $Cop$  & & $0.822\ (0.007)$ & $0.823\ (0.007)$ & $0.823\ (0.007)$ & & & $0.836\ (0.014)$ & $0.837\ (0.013)$ & $0.837\ (0.012)$ \\
                     & $Cox$     & & $0.819\ (0.007)$ & $0.823\ (0.007)$ & $0.822\ (0.007)$ & & & $0.838\ (0.013)$ & $0.838\ (0.013)$ & $0.838\ (0.012)$ \\
                                               \\
          \rowcolor{Gainsboro!60}
              & $Cop\ \star$   & & $0.816\ (0.007)$ & $0.816\ (0.006)$ & $0.815\ (0.006)$ &&& $0.863\ (0.014)$ & $0.863\ (0.012)$ & $0.863\ (0.013)$ \\
                & $Cox\ \star$ & & $0.816\ (0.007)$ & $0.816\ (0.006)$ & $0.815\ (0.006)$ &&& $0.864\ (0.014)$ & $0.863\ (0.012)$ & $0.863\ (0.013)$ \\
         \midrule
      \rowcolor{Gainsboro!60}
C-Index ($T_{2}$) & $Cop$  & & $0.855\ (0.005)$ & $0.855\ (0.006)$ & $0.855\ (0.006)$ & & & $0.948\ (0.005)$ & $0.948\ (0.006)$ & $0.947\ (0.006)$ \\
                       & $Cox$  & & $0.852\ (0.005)$ & $0.854\ (0.006)$ & $0.854\ (0.006)$ & & & $0.948\ (0.005)$ & $0.948\ (0.006)$ & $0.948\ (0.005)$ \\
                                               \\
    \rowcolor{Gainsboro!60}
                & $Cop\ \star$ & & $0.853\ (0.006)$ & $0.853\ (0.006)$ & $0.853\ (0.006)$ &&& $0.911\ (0.014)$ & $0.911\ (0.013)$ & $0.911\ (0.014)$ \\
                & $Cox\ \star$ & & $0.852\ (0.007)$ & $0.853\ (0.006)$ & $0.853\ (0.006)$ &&& $0.911\ (0.014)$ & $0.910\ (0.014)$ & $0.910\ (0.015)$ \\
     % $\ ()$ & $\ ()$ & $\ ()$ & & & $\ ()$ & $\ ()$ & $\ ()$ \\
      \midrule
      \midrule
                                %\rowcolor{Gainsboro!60}
    \multicolumn{11}{l}{Clayton copula with Kendall's $\tau$ with range within $[0.159;\ 0.907]$ in linear DGP, and  $[0.022;\ 0.917]$ in non-linear DGP. }  \\
    \\
    \multicolumn{11}{l}{Gradients stabilised using $L_2$ norm, step-length $\texttt{s}_{\texttt{step}} = 0.1$. $n_{\text{train}} = 1000$, $n_{\text{test}} = 1000$, $n_{\texttt{mstop}} = 1000$. } \\
\bottomrule
    \end{tabular}
    }\label{SimulationsAllMetrics_TOEP}
\end{table}
}

\pagebreak
\begin{table}[t!]
    \centering
    \caption{\small Simulation study 2. True positive rates (TPR) and false positive rates (FPR) for the copula ($Cop$) models for each distribution parameter, as well as independent univariate Cox models ($Cox$) for each margin, $\star$ denotes non-linear DGP. 
    Values are averages over $500$ independent datasets. 
    }
    \resizebox{\columnwidth}{!}{%
    \begin{tabular}{l cccccccc ccccccccccc } % 20 columns
    \toprule
           &   & \multicolumn{8}{c}{$(1)$} & & & \multicolumn{8}{c}{ $(2)$ }  \\
      \\
  & & \multicolumn{8}{c}{ 30\% censoring } & & &  \multicolumn{8}{c}{ 70\% censoring }  \\
  \\
 & & \multicolumn{2}{c}{ $p_1 = 10$ } & & \multicolumn{2}{c}{ $p_2 = 500$ } & & \multicolumn{2}{c}{ $p_3 = 1000$ } & & & 
     \multicolumn{2}{c}{ $p_1 = 10$ } & & \multicolumn{2}{c}{ $p_2 = 500$ } & & \multicolumn{2}{c}{ $p_3 = 1000$ }  \\
             \\
   &  &  TPR & FPR & & TPR & FPR & & TPR & FPR & & & TPR & FPR  & & TPR & FPR & & TPR & FPR \\
              %& \\  
         \midrule
         %  \multirow{15}{*}{\STAB{\rotatebox[origin=c]{90}{rota}}} \\
     & &    \multicolumn{18}{c}{Linear DGP}  \\
        % \midrule
   \multicolumn{13}{l}{Copula model $(Cop)$ }  \\
   \\
  \rowcolor{Gainsboro!60}
$\vartheta_{1}^{(1)}$ &  &  $1$ & $0.260$ &  & $1$ & $0.025$ &  & $1$ & $0.013$ &  &  & $1$ & $0.281$ &  & $1$ & $0.032$ &  & $1$ & $0.018$ \\
$\vartheta_{2}^{(1)}$ &  &  $1$ & $0.164$ &  & $1$ & $0.001$ &  & $1$ & $0.000$ &  &  & $1$  & $0.190$ &  & $1$ & $0.003$  &  & $1$ & $0.001$ \\
\\
 \rowcolor{Gainsboro!60}
$\vartheta_{1}^{(2)}$ &  & $1$ & $0.251$  &  & $1$ & $0.035$ &  & $1$ & $0.021$ &  &  & $1$ & $0.325$ &  & $1$ & $0.025$ &  & $1$ & $0.011$ \\
$\vartheta_{2}^{(2)}$ &  & $1$ & $0.164$  &  & $1$ & $0.002$ &  & $1$ & $0.001$ &  &  & $1$ & $0.190$ &  & $0.394$ & $0.000$ &  & $0.261$ & $0.000$ \\
\\
 \rowcolor{Gainsboro!60}
$\vartheta_{}^{(c)}$ &  & $1$ & $0.301$  &  & $1$ & $0.076$ &  & $1$ & $0.040$ &  &  & $0.993$ & $0.221$ &  & $0.698$ & $0.021$ &  & $0.604$ & $0.010$ \\
\midrule
 \multicolumn{13}{l}{Cox models $(Cox)$ }  \\ %%  UNIVARIATE MODEL!!!
 \\
 \rowcolor{Gainsboro!60}
Margin 1 &  & $0.855$ & $0.234$ &  & $0.528$ & $0.035$ &  &  $0.491$ & $0.025$  &&& $0.877$ & $0.193$ & & $0.799$ & $0.032$ & & $0.793$ & $0.020$\\
Margin 2 &  & $0.951$ & $0.210$ & & $0.888$ & $0.041$ & & $0.869$ & $0.028$     &&& $0.945$ & $0.195$ & & $0.741$ & $0.041$ & & $0.719$ & $0.024$ \\
\midrule
\midrule
& & \multicolumn{18}{c}{Non-linear DGP}  \\
%%%%%%%%% NON-LINEAR SETTINGS
%\midrule
 \multicolumn{13}{l}{Copula model $(Cop\ \star)$ }  \\
 \\
  \rowcolor{Gainsboro!60}
$\vartheta_{1}^{(1)}$ &  & $1$ & $0.190$ & & $1$ & $0.011$ & & $1$ & $0.006$ &&& $1$ & $0.214$ & & $1$ & $0.012$ &  & $1$ & $0.006$ \\
$\vartheta_{2}^{(1)}$ &  & $1$ & $0.227$ & & $1$ & $0.011$ & & $1$ & $0.006$ &&& $1$ & $0.244$ & & $1$ & $0.013$ &  & $1$ & $0.007$ \\
\\
 \rowcolor{Gainsboro!60}
$\vartheta_{1}^{(2)}$ &  & $1$ & $0.264$  & & $1$  & $0.019$ & & $1$ & $0.009$  &&& $1$ & $0.269$ & & $1$ & $0.016$ &  & $1$ & $0.008$ \\
$\vartheta_{2}^{(2)}$ &  & $1$ & $0.227$  & & $0.978$ & $0.005$ &  & $0.960$ & $0.003$  &&& $1$ & $0.244$ & & $0.696$ & $0.004$ &  & $0.576$ & $0.002$ \\
\\
 \rowcolor{Gainsboro!60}
$\vartheta_{}^{(c)}$ &  & $1$  &  $0.271$ & & $1$ & $0.016$ & & $1$ & $0.008$ &&& $1$ & $0.132$ & & $0.994$ & $0.004$ &  & $0.998$ & $0.001$ \\
\midrule
 \multicolumn{13}{l}{Independent univariate Cox models $(Cox\ \star)$ }  \\%%  UNIVARIATE MODEL!!!
 \\
 \rowcolor{Gainsboro!60}
Margin 1 &  & $0.759$  & $0.253$ & & $0.533$ & $0.042$ & & $0.532$ & $0.029$ &&& $0.893$ & $0.257$ & & $0.688$ & $0.039$ &  & $0.662$ & $0.026$ \\
Margin 2 &  & $0.897$  & $0.246$ & & $0.687$ & $0.052$ & & $0.641$ & $0.034$ &&& $0.834$ & $0.280$ & & $0.553$ & $0.051$ &  & $0.543$ & $0.033$ \\
\midrule
\midrule
  \multicolumn{20}{l}{Clayton copula with Kendall's $\tau$ with range within $[0.159;\ 0.907]$ in linear DGP, and  $[0.022;\ 0.917]$ in non-linear DGP. }  \\
    \\
    \multicolumn{20}{l}{Gradients stabilised using $L_2$ norm, step-length $\texttt{s}_{\texttt{step}} = 0.1$. $n_{\text{train}} = 1000$, $n_{\text{test}} = 1000$, $n_{\texttt{mstop}} = 1000$. } \\
\bottomrule
    \end{tabular}
    }
    \label{Simulation1SelectionRates_TOEP}
\end{table}

\end{document}

%% file: defs.tex
\def \dsP {\text{$\mathds{P}$}}

\def \dsR {\text{$\mathds{R}$}}

\DeclareMathOperator*{\argmin}{{arg\,min}}

\def \xvec {\text{\boldmath$x$}}    
\def \yvec {\text{\boldmath$y$}}    \def \mY {\text{\boldmath$Y$}}

\def \deltavec        {\text{\boldmath$\delta$}}

\def \varthetavec     {\text{\boldmath$\vartheta$}}